\documentclass[showpacs,amsmath,amssymb,preprint]{revtex4}
\usepackage{here}
\usepackage{setspace}
\usepackage{graphicx}
\usepackage{dcolumn}
\usepackage{bm}
\usepackage{color}
\begin{document}


\title{Thermalization of Atom-Molecule Bose Gases in a Double-Well Potential}

\author{Atsushi Motohashi}
\email{motohashi2011@gmail.com}
\affiliation{%
Department of Physics, Tokyo University of Science,
 1-3 Kagurazaka, Shinjuku-ku, Tokyo 162-8601, Japan\\
}%

\date{\today}

\begin{abstract}
We study the non-equiliribium dynamics of atom-molecule Bose gases in a double-well potential. In this system, the internal atom-molecule tunneling has significant influence on the dynamics. We investigate the periodicity of dynamics by studying the level statistics of the quantum system. We find that chaotic energy eigenstates arise from the competition between the interwell and the atom-molecule internal tunnelings. Furthermore, we show that the physical quantities relax to the microcanonical averages in the full-quantum dynamics when the system is chaotic. This thermalization is caused by the verification of eigenstate thermalization hypothesis (ETH). We show numerically that the onset of ETH occurs simultaneously with that of chaos. In addition, we show that the energy eigenstates become to be exponentially localized states simultaneously with the onset of chaos.
\end{abstract}

\pacs{Valid PACS appear here}
\maketitle

\section{Introduction}

Recently, the experimental studies of quantum signature of chaos have been achieved in ultracold atomic systems \cite{KR_Exp_1995,KR_2008,KR_2009_PRA,nature_chaos}. For instance, the dynamical localization, which is the quantum counterpart of chaos in classical systems \cite{KR_1984,Garcia_PRL_2008}, has been observed. The quantum signature of chaos has become one of the important subjects in cold atom physics. 

In this paper, we study the non-equiliribium quantum dynamics of atom-molecule Bose gases in a double-well potential. Recently, the tunneling dynamics between left and right wells is observed in Bose-Einstein condensates trapped in a double-well potential \cite{BJJ}. This experimental achievement has triggered much interesting research \cite{Chaiotic_BJJ,Shchesnovich_Trippenbach,Dissipated_BJJ}. Though the spatial coherence of Bose gases is focused in these studies, tunneling effects occur not only between spatially separated states but also between internal degrees of freedom. In particular, the effects of internal tunneling between atomic and molecular states have been discussed theoretically \cite{Rarified_Liquid_AM_BEC,Guang,Santos}. Furthermore, in an atom-molecule mixture in an optical lattice, the atom-molecule internal tunneling dynamics has been observed \cite{AM_BEC_Wieman,Ryu,AMRabi}. Based on these studies, we study atom-molecule Bose gases in a double-well potential with a particular focus on the effect of atom-molecule internal tunneling. As a result, we show that the atom-molecule internal tunneling induces chaos and thermalization dynamics.

In general full-quantum systems, the chaotic dynamics can not occur because the Schr{\"o}dinger equation is the linear equation (see Ref. \cite{Saito_Makino} and references there in). The quantum signature of the semiclassical non-periodicity is an important problem \cite{Berry_1}. In the preceding studies \cite{BGS,Shudo_Saito}, it is conjectured that the quantum signature of chaos appears in the level spacing distribution. Based on these considerations, we investigate the statistical property of energy spectra. As a result, we conclude that the atom-molecule internal tunneling induces the chaotic energy eigenstates.

Furthermore, we discuss the relation between chaotic energy eigenstates and thermalization. Recently, the experimental challenge for the foundation of statistical mechanics has become possible. By using quasi-one-dimensional Bose gases, the non-equilibrium dynamics of isolated quantum many-body systems is investigated experimentally \cite{Kinoshita,Hofferberth}. From 40 to 250 $^{87}\textrm{Rb}$ atoms do not thermalize \cite{Kinoshita}. It is showed that the multiplicity of degrees of freedom induces thermalization generally \cite{Sugita_2007}. In the experiment \cite{Kinoshita}, it seems that the number of degrees of freedom is not large enough for thermalization. 

However some numerical studies exhibit that the non-integrability of systems induced thermalization even in a few body systems \cite{Rigol_Nature,Kollath_PRL_2007,Rigol_PRL_2007,Rigol_PRL_2009,Rigol_gapped,Roux,Kollath_Roux}. In these studies, the integrability of the Bose-Hubbard model is estimated by investigating the level spacing statistics. When the system is non-integrable, the semi-classical dynamics is chaotic, otherwise the system exhibits regular dynamics. The preceding studies investigated the quantum dynamics in the non-integrable region. The numerical results indicate that when the semi-classical dynamics is chaotic, thermalization occurs in the corresponding quantum dynamics. Otherwise thermalization does not occur. The integrability of systems plays an important role in the thermalization mechanism of isolated-quantum systems. 

In addition, the system size is also important for thermalization. The preceding studies focused on single-component Bose gases in a optical lattice. These studies showed the appearance of non-integrability by investigating the Bose gasses confined in about 20 sites \cite{Rigol_Nature,Rigol_gapped} and 8$\sim$12 sites \cite{Kollath_Roux} optical lattices. In particular, Kollath {\it et al.}\cite{Kollath_Roux} showed that the non-integrability is increased by increasing the system size. From this result, the system size should be large to some extent in order for thermalization to occur. Furthermore, in the preceding studies, in the region where the interwell tunneling and interparticle interaction compete, the system exhibits thermalization. 

In contrast to these studies, we show that thermalization can occur in the system with only two sites. Our subject is atom-molecule Bose gases in a double-well potential. In this system, the internal atom-molecule tunneling induces chaos. In addition, we confirm that thermalization occurs in the corresponding quantum system. The large system size is not always needed for thermalization. Furthermore, this thermalization process dose not need interparticle interactions. The competition between the interwell tunneling and the atom-molecule internal tunneling induces the thermalization dynamics. 

We also find that the energy eigenstates become to be exponentially localized. This is quite similar to Anderson localization. In the previous studies, it is pointed out the relation between quantum chaos and the Anderson localization in Kicked rotor\cite{KR_1984}. Furthermore, this is confirmed clearly in the experiments\cite{KR_2008,KR_2009_PRA}. However, this relation is not confirmed in general cases. In this paper, we investigate whether the exponential localization occurs in our system, and find it does indeed.  

\section{Model}
\subsection{Hamiltonian}
The second-quantized Hamiltonian for Bose atoms and molecules can be written as
\begin{eqnarray}
&& \hat{H} = \sum_{i=a,b} \int d \mathbf{r} \Bigg( \frac{ \hbar^2 }{ 2m_{i} } \nabla \hat{ \Psi }_{i}^{ \dag } \cdot \nabla \hat{ \Psi }_{i} + V_{ \rm{ext} } ( \mathbf{ r }  )  \hat{ \Psi }_{i}^{ \dag } \hat{ \Psi }_{i} \Bigg) + \frac{ g_{a} }{2} \int d {\bf r} \hat{ \Psi }_{a}^{ \dag } \hat{ \Psi }_{a}^{ \dag } \hat{ \Psi }_{a} \hat{ \Psi }_{a} \nonumber \\
&& \qquad - \lambda \int d \mathbf{r} \left( \hat{ \Psi }_{b}^{ \dag } \hat{ \Psi }_{a} \hat{ \Psi }_{a} + \hat{ \Psi }_{a}^{\dag}  \hat{ \Psi }_{a}^{\dag} \hat{ \Psi }_{b} \right) + \delta \int d \mathbf{r} \hat{ \Psi }_{b}^{ \dag } \hat{ \Psi }_{b} ,  \qquad 
\label{eq:1}
\end{eqnarray}
where $\hat{ \Psi }_{a}$ and $\hat{ \Psi }_{b}$ represent  field operators for Bose atoms and molecules respectively, $\lambda$ is the internal tunneling strength between atomic and molecular states, $\delta$ is the energy difference between atoms and molecules, and $V_{ \rm{ext} } ( \mathbf{ r } )$ is a double-well potential. $m_{a}$ is the atomic mass. $m_{b} = 2 m_{a}$ is the molecular mass. The interatomic interaction can be approximated in terms of the $s$-wave scattering lengths as $g_{a} = 4 \pi \hbar^2 a_{s} / m_{a}$. The intermolecule, and the atom-molecule interactions are neglected in our treatment. Furthermore, we introduce the four-mode approximation. In this approximation, we concentrate on only lowest-energy atomic and molecular modes in each well, and ignore the effect of the particles occupying other modes. From this point of view, field operators can be approximated as $\hat{ \Psi }_{a} \simeq \Phi_{aL} \hat{a}_{L} + \Phi_{aR} \hat{a}_{R}, \hat{ \Psi }_{b} \simeq \Phi_{bL} \hat{b}_{L} + \Phi_{bR} \hat{b}_{R}$, where $\Phi_{aL}, \Phi_{aR}$ ($\Phi_{bL}, \Phi_{bR}$) are the wavefunctions of the atomic (molecular) lowest-energy modes in the left well and the right well respectively. $\hat{a}_{L}, \hat{a}_{R}$($\hat{b}_{L}, \hat{b}_{R}$) are annihilation operators for the atomic (molecular) lowest-energy modes in the left well and the right well respectively. Applying these approximations to Eq. (\ref{eq:1}), we obtain the quantum four-mode Hamiltonian (four-mode model). 
\begin{eqnarray}
&&\hat{H} = - J_{a} \big( a_{L}^{\dag} a_{R} +  a_{R}^{\dag} a_{L} \big) - J_{b} \big( b_{L}^{\dag} b_{R} + b_{R}^{\dag} b_{L} \big) + \Delta \big( b_{L}^{\dag} b_{L} + b_{R}^{\dag} b_{R} \big)  \nonumber \\
&& \qquad + \frac{ U_{a} }{2} \big( a_{L}^{\dag} a_{L}^{\dag} a_{L} a_{L} + a_{R}^{\dag} a_{R}^{\dag} a_{R} a_{R} \big) \nonumber \\
&& \qquad - g \big( b_{L}^{\dag} a_{L} a_{L} + b_{R}^{\dag} a_{R} a_{R} + a_{L}^{\dag} a_{L}^{\dag} b_{L} + a_{R}^{\dag} a_{R}^{\dag} b_{R} \big) , \quad
\label{eq:am_H_2}
\end{eqnarray}
where the parameters are defined in Appendix. In order to study the non-equilibrium dynamics, we obtain all the eigenenergies and eigenvectors by performing the full exact diagonalization of the Hamiltonian (\ref{eq:am_H_2}).

In what follows, we write the $i$th energy eigenstate as 
\begin{eqnarray}
| \Psi_{i} \rangle = \sum_{N_{aL}, N_{aR}, N_{bL}, N_{bR}} \Phi_{i} \left( N_{aL}, N_{aR}, N_{bL}, N_{bR} \right) | N_{aL}, N_{aR}, N_{bL}, N_{bR} \rangle,
\label{ith_eigen}
\end{eqnarray}
where $N_{aL(bL)}$ and $N_{bL(bR)}$ are the atomic (molecular) particle numbers in the left and right wells. $| N_{aL}, N_{aR}, N_{bL}, N_{bR} \rangle$ is the eigenstate of the operators $\hat{a}^{\dag}_{L} \hat{a}_{L}$, $\hat{a}^{\dag}_{R} \hat{a}_{R}$, $\hat{b}^{\dag}_{L} \hat{b}_{L}$ and $\hat{b}^{\dag}_{R} \hat{b}_{R}$. The total particle number $N \equiv N_{aL} + N_{aR} + 2 \left( N_{bL} + N_{bR} \right)$ is always conserved. In this paper, we calculate the summation $\sum_{N_{aL}, N_{aR}, N_{bL}, N_{bR}}$ under the condition of the conservation of the total particle number $N$. We write the expectation value of the $i$th energy eigenstate as $\langle \hat{ A } \rangle_{ i } \equiv \langle \Psi_{ i } | \hat{ A } | \Psi_{ i } \rangle$. $\hat{A}$ is any observable.

We set the parameters in the Hamiltonian. We set the molecular-tunneling strength as $J_{b} / J_{a} = \frac{1}{2}$. Next, we consider the atom-molecule energy difference $\Delta$. This is controllable parameter in the experiment by means of Feshbach resonances. In this study, we set this value as $\Delta / J_{a} = - 1$. We set $U_{a} = 0$ except Sec. \ref{sec:Ua}. In only Sec. \ref{sec:Ua}, we set $U_{a} \ne 0$ and investigate the influence of the interparticle interaction on our results.

\subsection{Time-evolution of particle numbers}
The initial state is expanded as $| \Psi \left( t = 0 \right) \rangle = \sum_{\alpha} C_{\alpha} | \Psi_{\alpha} \rangle$, where $| \Psi_{\alpha} \rangle$ represents the $\alpha$th energy eigenstate. The many-body wavefunction evolves as $| \Psi \left( t \right) \rangle = e^{ - i \hat{ H } t / \hbar } | \Psi \left( t = 0 \right) \rangle = \sum_{\alpha} C_{\alpha} e^{ - i E_{\alpha} t } | \Psi_{\alpha} \rangle$, where $E_{\alpha}$ is the $\alpha$'s eigenenergies. The quantum-mechanial expectation of any observable $\hat{A}$ evolves as  
\begin{eqnarray}
\langle \hat{A} \rangle = \langle \Psi \left( t \right) | \hat{A} | \Psi \left( t \right) \rangle = \sum_{\alpha, \beta} C_{\alpha}^{*} C_{\beta} e^{ - i \left( E_{ \alpha } - E_{ \beta } \right) t } A_{ \alpha \beta },
\end{eqnarray}
where $A_{\alpha \beta} \equiv \langle \Psi_{\alpha} | \hat{ A } | \Psi_{\beta} \rangle$.  The long-time average of $\langle \hat{A} \rangle$ is 
\begin{eqnarray}
\overline{ \langle \hat{ A } \rangle } = \sum_{\alpha} | C_{\alpha} |^2 A_{\alpha \alpha}.
\label{eq:relax}
\end{eqnarray}
If the system relaxes to equilibration, this is the relaxed value of the physical quantities. In this paper, we investigate whether the physical quantities relax to the microcanonical ensemble average. The microcanonical ensemble average is given as
\begin{eqnarray}
\langle \hat{A} \rangle^{micro}_{i} \equiv \frac{ 1 }{ 2 \Delta N + 1 } \sum_{i^{'}=i - \Delta N, i + \Delta N} \langle \hat{A} \rangle_{ i^{'} }, \label{eq:micro_canonical_mean_value}
\end{eqnarray}
where the $i$th energy eigenstate has the closest energy eigenvalue to the mean energy of the initial state. We take the $2 \times \Delta N$ energy eigenstates around the $i^{'}$th energy eigenstate, and calculate the microcanonical average using the principle of equal weight. Unless $\Delta N$ is not too large, any other values of $\Delta N$ does not change our results as explained in Sec. \ref{sec:ETH}.  

\section{Level statistics}
In this section, we show that the atom-molecule internal tunneling induces the chaotic energy eigenstates. First, we introduce the indicator for the quantum signature reflecting chaos in corresponding semi-classical systems. The quantum signature of chaos appears in level spacing distributions. 

Here we explain how the author calculates level spacing distribution. First we perform the exact diagonalization of the Hamiltonian matrix (\ref{eq:am_H_2}) and obtain the eigenenergies $E_{i}$ ($i = 1, ..., N_{T}$). $N_{T}$ is the number of the eigenstates. $E_{i}$ are arranged in ascending order. Second we calculate the adjacent level spacings as $S_{i} = E_{i+1} - E_{i}$ ($i = 1, ..., N_{T} - 1$). We define the average of the adjacent level spacings as $\bar{S} \equiv \sum_{i=1}^{ N_{T} - 1} S_{i} / \left( N_{T} - 1 \right)$. Next we divide the ($E_{ N_{T} } - E_{1}$) by $0.16 \times \bar{S}$. Then we count the number of eigenstates included in the intervals $\Delta S_{i} \equiv \left[ \left( i - 1 \right) \times 0.16 \bar{S}, i \times 0.16 \bar{S} \right]$. Finally we write the histogram. The horizontal axis is normalized by $\bar{S}$, and the vertical axis is normalized so as to $\int d S P \left( S \right) = 1$. We note that we can change $0.16$. Our results are not so sensitive to this value. 

In a quantum mechanical system, the dynamics is always periodic and chaos does not occur, because the time-evoluiton equation is linear. However, the breakdown of periodicity in the classical counterpart has influence on the level spacing distribution of the quantum system, because the level spacing is related with the dynamical nature of physical quantities as $\langle \hat{A} \rangle =  \sum_{\alpha, \beta} C_{\alpha}^{*} C_{\beta} e^{ - i \left( E_{ \alpha } - E_{ \beta } \right) t } A_{ \alpha \beta }$. Intuitively, level repulsions correspond to chaos. When chaos occurs, many periods are included in the dynamics in the semiclassical case. Therefore, the transition between energy eigenstates occur in the quantum counterpart, and level repulsions are induced. In particular it has been conjectured that, when chaos occurs in the classical counterpart, the level spacing distribution in the quantum system shows the universal features, which are related to the random matrix theory.  \cite{BGS} 

\begin{figure}
\begin{minipage}{18pc}
\includegraphics[ width=0.9\linewidth, keepaspectratio]{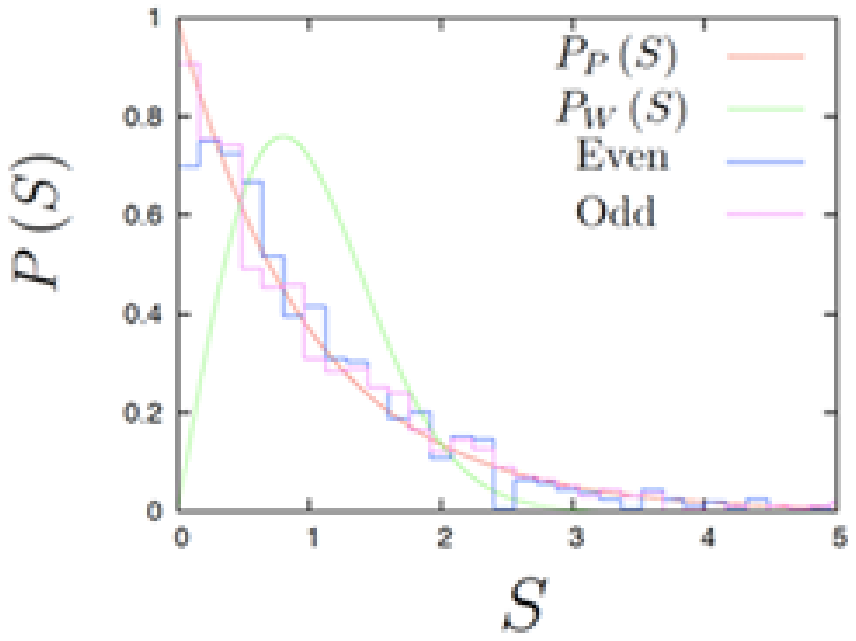}
\caption{ (color online) Level spacing distributions at $\sqrt{N}g/J_{a}=0.5$. $N=40$. Blue and violet bars represent odd and even parity energy spectra. Red and green lines are Poisson and Wigner distributions. The lower and higher 20 \% levels are not included.}
\label{LS_g_low}
\end{minipage}\hspace{1.0pc}
\begin{minipage}{18pc}
\includegraphics[ width=0.9\linewidth, keepaspectratio]{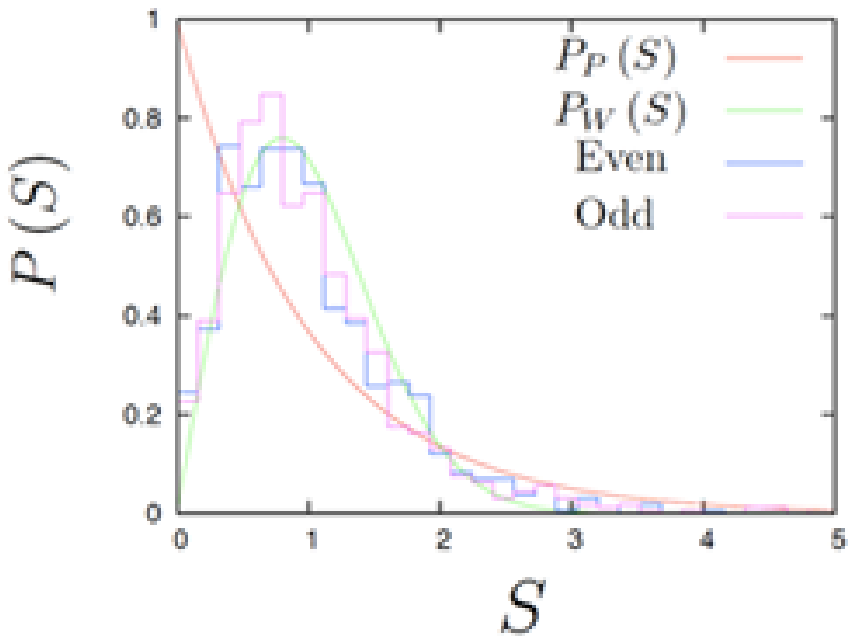}
\caption{ (color online) Level spacing distributions at $\sqrt{N}g/J_{a}=1.5$ and $N=40$. Blue and violet bars represent odd and even parity energy spectra. Red and green lines are Poisson and Wigner distributions. Lower 20 \% and higher 20 \% levels are omitted.}
\label{LS_g_mid}
\end{minipage}
\end{figure}

From the random matrix theory, the level-spacing distribution of a spinless and time-revaersal Hamiltonian exhibits the Wigner-Dyson distribution 
\begin{eqnarray}
P_{W} \left( S \right) = \frac{ \pi S }{ 2 } \exp{ \left( - \pi S^{2} / 4 \right) },
\end{eqnarray}
where $S$ is the adjacent level spacing normalized by the mean-level spacing. $P_{W} \left( S \right)$ takes the maximum value at $S \ne 0$. This indicates that level repulsions occur. When the dynamics is chaotic in classical systems, it is conjectured that the level-spacing distribution of the quantum counterpart is the Wigner-Dyson distribution \cite{BGS}. In contrast to chaotic cases, when the system is integrable, the level-spacing distribution is the Poissonian distribution \cite{Berry_2}
\begin{eqnarray}
P_{P} \left( S \right) = e^{- S}.
\end{eqnarray}
If the level-spacing distribution follows the Poissonian distribution, the semi-classical dynamics is regular. In contrast to chaotic cases, $P_{P} \left( S \right)$ has the maximum value at $S = 0$. This indicates that level clustering occurs.

Before the analysis, we need to classify the energy spectra according to the symmetry of the system. The symmetry has significant influence on level repulsions \cite{Chaos_OL}. Since the double-well potential has the left-right symmetry, the energy eigenstates can be classified into odd and even parity states. By interchanging the left and right well particle numbers of the eigenfunctions as $N_{aL} \leftrightarrow N_{aR}$ and $N_{bL} \leftrightarrow N_{bR}$, even parity states are not changed, while the sign of odd parity states are changed as $| \Psi \rangle \rightarrow - | \Psi \rangle$. We should investigate the odd and even parity spectra separately. In Figs. \ref{LS_g_low} and \ref{LS_g_mid}, we show the level spacing distribution of the even and odd parity spectra. $S$ is normalized by the mean adjacent level spacing $\bar{S}$. We do not include the lower 20 \% and higher 20 \% levels because these eigenstates are always regular, not chaotic.

When the atom-molecule internal tunneling is small $(\sqrt{ N } g / J_{a} = 0.5)$, the level spacing distributions are close to the Poisson distribution. However, the level spacing distributions come close to the Wigner distribution in the region where the atom-molecule internal tunneling and the interwell tunneling compete $(\sqrt{ N } g / J_{a} = 1.5)$. The dynamical property is significantly changed by the atom-molecule internal tunneling. In the next section, we investigate the time evolutions of the physical quantities in these regions. 

\begin{figure}
\begin{center}
\includegraphics[width=12.3cm]{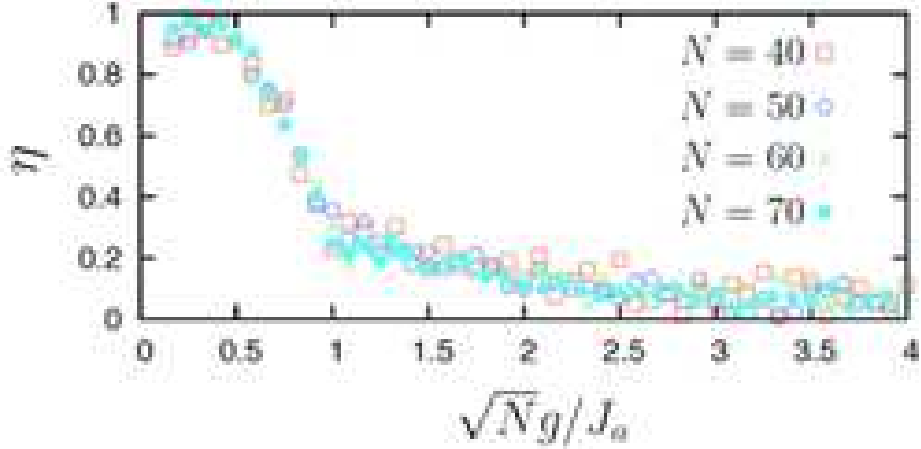}
\caption{ (color online) The $g$-dependence of $\eta$. In this figure, only even spectra are included. The lower and higher 20 \% levels are not included.}
\label{fig:eta}
\end{center}
\end{figure}

In order to evaluate the periodicity of semiclassical dynamics quantitatively, we define the indicator
\begin{eqnarray}
\eta \equiv \left| \frac{ \int_{0}^{S_{0}} d S \left[ P \left( S \right) - P_{W} (S) \right] }{ \int_{0}^{S_{0}} d S \left[ P_{P} \left( S \right) - P_{W} (S) \right]  } \right|,
\label{eq:eta}
\end{eqnarray}
where $P_{W} \left( S \right)$ and $P_{P} \left( S \right)$ intersect at $S_{0} \simeq 0.473$. When $\eta = 1$, $P \left( S \right) = P_{P} \left( S \right)$ and the dynamics is periodic. When $\eta = 0$, $P \left( S \right) = P_{W} \left( S \right)$ and the dynamics is chaotic. In Fig. \ref{fig:eta}, we investigate $\eta$ by increasing the atom-molecule internal tunneling strength $g$. In this figure, we use the even spectra and do not include the lower and higher $20 \%$ energy levels, because these energy-eigenstates are always regular. From this figure, we conclude that the atom-molecule internal tunneling changes the dynamics drastically. Chaos is induced by the atom-molecule internal tunneling. 

Here, we note that our results in Fig. \ref{fig:eta} is independent of the interwell molecular tunneling $J_{b}$.  In the Wigner region ($\sqrt{N} g / J_{a} = 4$), we investigate $\eta$ by changing $J_{b}$ from 0 to 1. As a result, $\eta$ is almost insensitive to $J_{b}$. Chaos is induced by the internal atom-molecule tunneling independently to the molecular interwell tunneling strength. 

\begin{figure}
\begin{minipage}{18pc}
\includegraphics[width=0.9\linewidth, keepaspectratio]{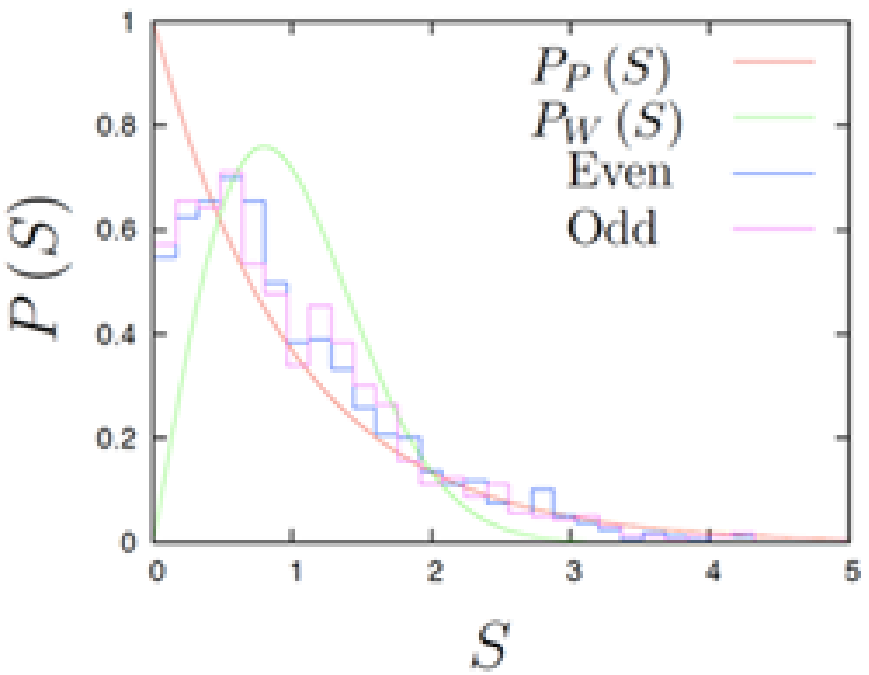}
\caption{ (color online) Level spacing distributions at $\sqrt{N}g/J_{a}=30$ and $N=40$. Blue and violet bars represent odd and even parity energy spectra. Red and green lines are Poisson and Wigner distributions.}
\label{fig:LS_g_high}
\end{minipage}\hspace{1.0pc}
\begin{minipage}{18pc}
\includegraphics[width=0.9\linewidth, keepaspectratio]{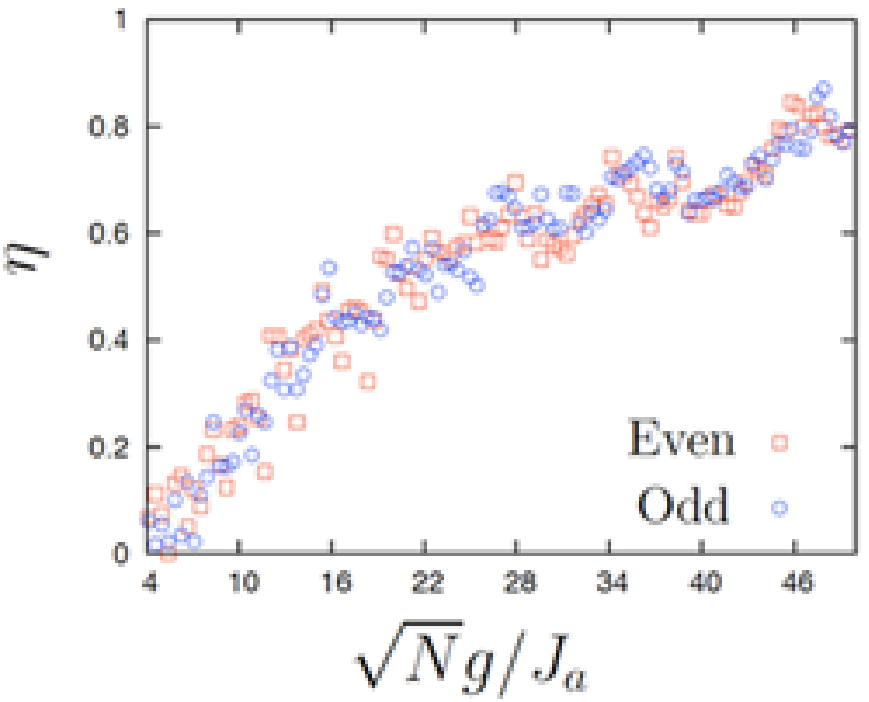}
\caption{ (color online) The $g$-dependence of $\eta$. $N = 40$. The lower and higher 20 \% levels are not included.}
\label{fig:eta_long}
\end{minipage}
\end{figure}

Furthermore, in Fig. \ref{fig:eta}, we found the scaling law where $\eta$ depends on the total particle number $N$ with the combination   $\sqrt{N} g / J_{a}$ . This indicates that the quantum fluctuation is weak as explained below. In the semi-classical limit, the creation-annihilation operators are replaced by c-numbers as $\hat{a}_{L(R)} \simeq \sqrt{N_{aL(aR)}} e^{i \theta_{ aL(aR) } }$, $\hat{b}_{L(R)} \simeq \sqrt{N_{bL(bR)}} e^{i \theta_{ bL(bR) } }$, and we obtain the semi-classical Hamiltonian as 
\begin{eqnarray}
H_{cl} / N &=& - 2 J_{a} \sqrt{ x_{L} x_{R} } \cos \left( \theta_{aR} - \theta_{aL} \right) - 2 J_{b} \sqrt{ y_{L} y_{R} } \cos \left( \theta_{bL} - \theta_{bR} \right) \nonumber \\
&& + \Delta \left( y_{L} + y_{R} \right) + \frac{ N U_{a} }{ 2 } \left( x_{L}^2 + x_{R}^2 \right) \nonumber \\
&& - 2 \sqrt{N} g \left[ x_{L} \sqrt{ y_{L} } \cos \left( 2 \theta_{aL} - \theta_{bL} \right) + x_{R} \sqrt{ y_{R} } \cos \left( 2 \theta_{aR} - \theta_{bR} \right) \right] ,
 \label{eq:classical_Hamiltonian_1}
\end{eqnarray}  
where the normalized particle numbers are defined as $x_{L(R)} \equiv N_{aL(aR)} / N$ and $y_{L(R)} \equiv N_{bL(bR)} / N$. In the classical Hamiltonian (\ref{eq:classical_Hamiltonian_1}), the atom-molecule internal tunneling strength $g$ is scaled by $\sqrt{N}$. Therefore, when the quantum fluctuation is weak, the physical quantities depend on $g$ via combination $\sqrt{N} g$. Then we conclude that the system behaves classically in a certain sense, although we perform the full-quantum analysis by using the full-exact diagonalization method. 

Next, we investigate the large $g$ region. In Fig. \ref{fig:LS_g_high}, we show the level spacing distributions at $\sqrt{N} g / J_{a} = 30$. The level spacing distributions are close to the Poisson distribution. In Fig. \ref{fig:eta_long}, we investigate the $g$-dependence of $\eta$ in the large $g$ region. $\eta$ comes close to unity in this region. The high atom-molecule internal tunneling recovers the periodicity of dynamics. 

In the region $\sqrt{N} g / J_{a} \gg 1$ ($\sqrt{ N } g / J_{a} \ll 1$) regions, the system becomes two independent interwell (internal) boson Josephson junctions. In a boson Josephson junction, the dynamical variables are the relative particle number and the relative phase. The number of degrees of freedom is two.

In order for chaos to occur, the dimension of the dynamical system must be greater than two. Since the trajectories in the phase space cannot intersect with each other, significantly complex trajectories cannot exist in 2D systems. However, in higher dimensional systems, the trajectories in the phase space travel around much more freely than 2D dynamical systems. 

In our system, chaos can occur in the region where the internal and interwell tunneling compete. We note that this is the necessary condition but not the sufficient condition for chaos. As shown in Fig. \ref{fig:eta}, chaos does not occur at $\sqrt{N} g / J_{a} = 0.4$, although the interwell and internal tunneling strengths compete.  

\section{Thermalization}
\subsection{relaxation dynamics}
\label{sec:TD}
In this section, we investigate the relaxation dynamics in both the Wigner region ($\sqrt{N} g / J_{a} = 4$) and Poisson region ($\sqrt{N} g / J_{a} = 0.2$). In particular, we focus on the dynamics of the atomic (molecular)-particle numbers in the left and right wells. 

\begin{figure}
\begin{center}
\includegraphics[width=10.3cm]{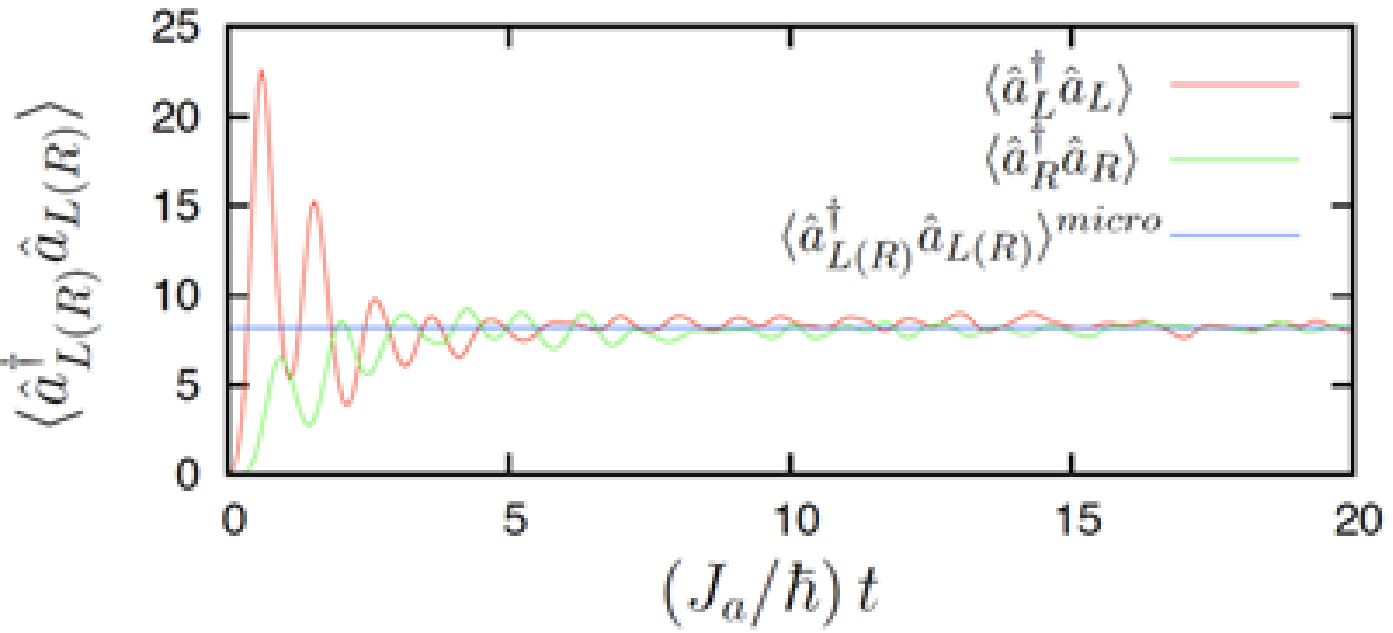}
\caption{ (color online) The time evolution of atomic particle numbers from the {\it case1} initial state. We investigate the Wigner region ($\sqrt{N} g / J_{a} = 4$).}
\label{fig:Time_evo_1_a}
\end{center}
\begin{center}
\includegraphics[width=10.3cm]{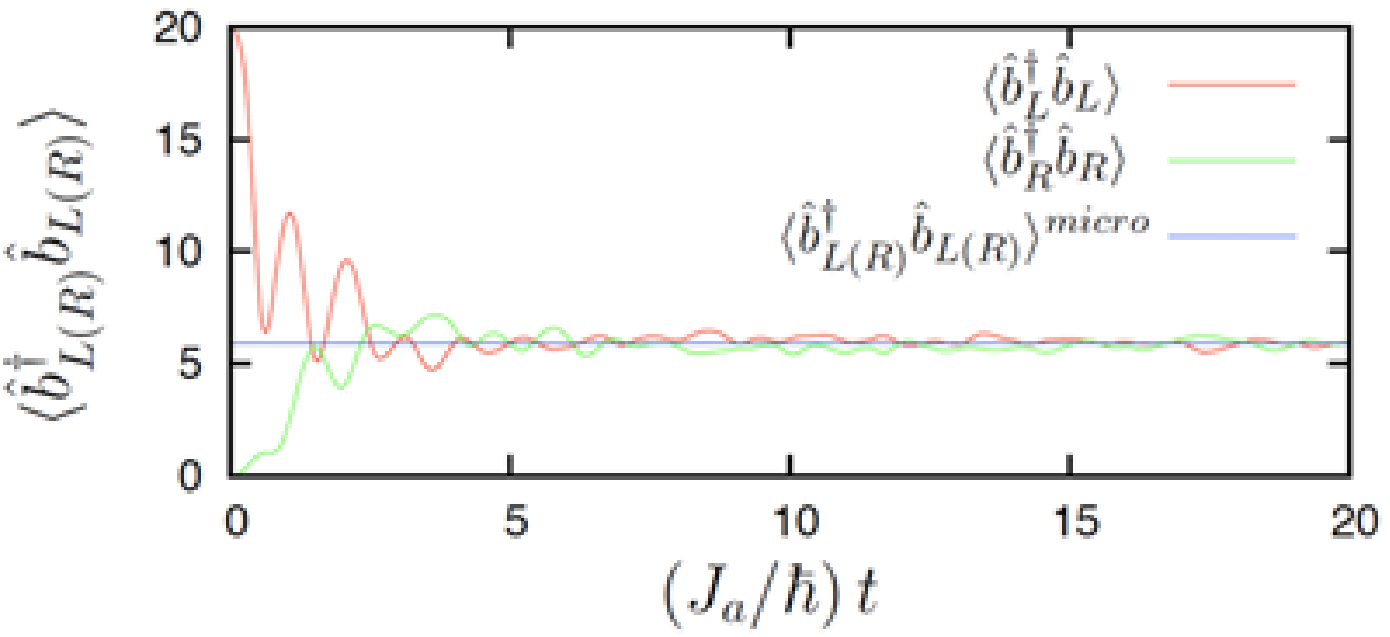}
\caption{ (color online) The time evolution of molecular particle numbers from the {\it case1} initial state. We investigate the Wigner region ($\sqrt{N} g / J_{a} = 4$).}
\label{fig:Time_evo_1_b}
\end{center}
\end{figure}

\begin{figure}
\begin{center}
\includegraphics[width=10.3cm]{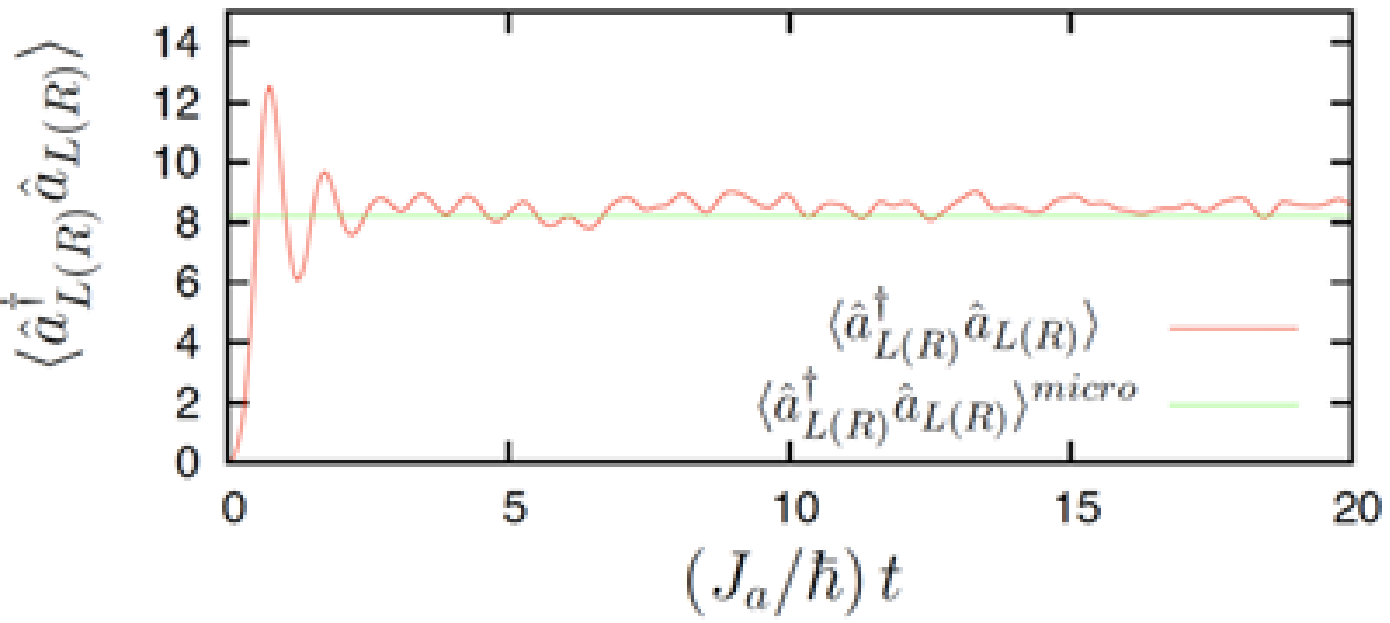}
\caption{ (color online) The time evolution of atomic particle numbers from the {\it case2} initial state. We investigate the Wigner region ($\sqrt{N} g / J_{a} = 4$). Red and green lines overlap.}
\label{fig:Time_evo_2_a}
\end{center}
\begin{center}
\includegraphics[width=10.3cm]{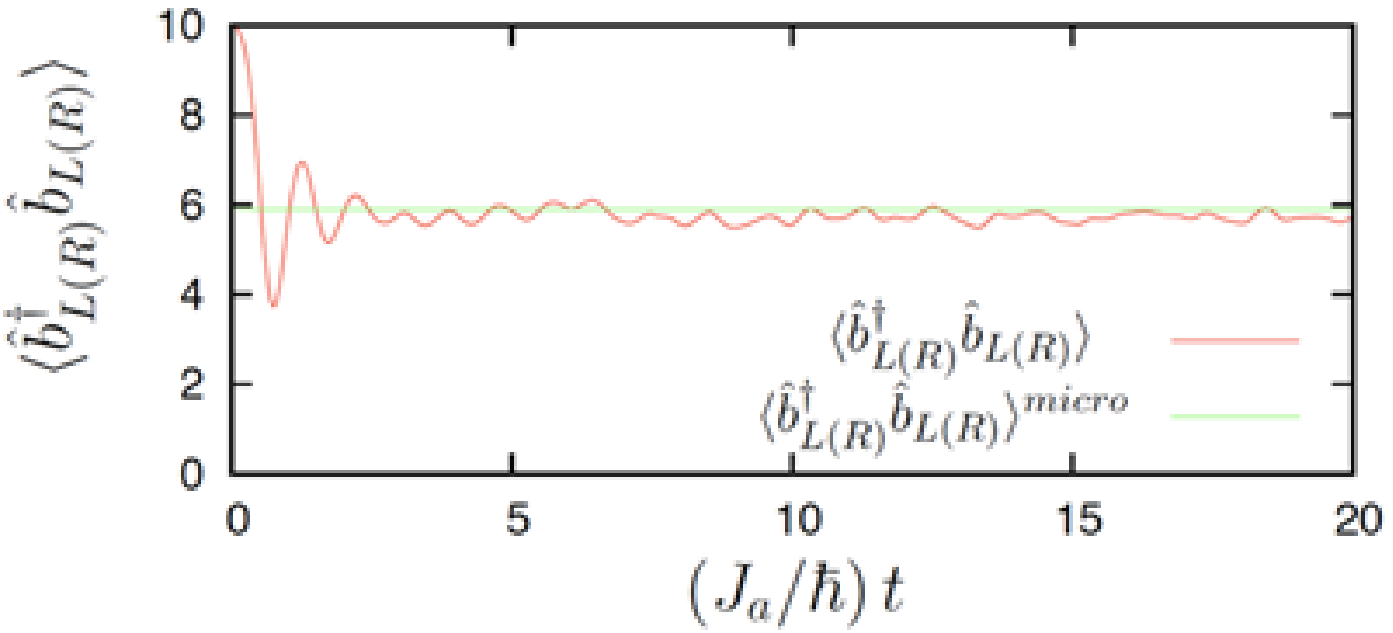}
\caption{ (color online) The time evolution of molecular particle numbers from the {\it case2} initial state. We investigate the Wigner region ($\sqrt{N} g / J_{a} = 4$). Red and green lines overlap.}
\label{fig:Time_evo_2_b}
\end{center}
\end{figure}

The important nature of thermal equilibration is that the final equilibrium state does not depend on the initial state. To confirm this, we investigate the relaxation from two different initial states, which have the same total energy. We call these different initial states {\it case1} and {\it case2}. In the initial state of {\it case1}, total particles localize in the left-well-molecular state only as $| 0, 0, N/2, 0 \rangle$. In the initial state of {\it case2}, a half of particles localizes in the left-well-molecular state, and another does in the right-well-molecular state as $| 0, 0, N/4, N/4 \rangle$. In these initial states, the total energy is $N \Delta$ from (\ref{eq:am_H_2}).

\begin{figure}
\begin{minipage}{18pc}
\includegraphics[width=0.9\linewidth, keepaspectratio]{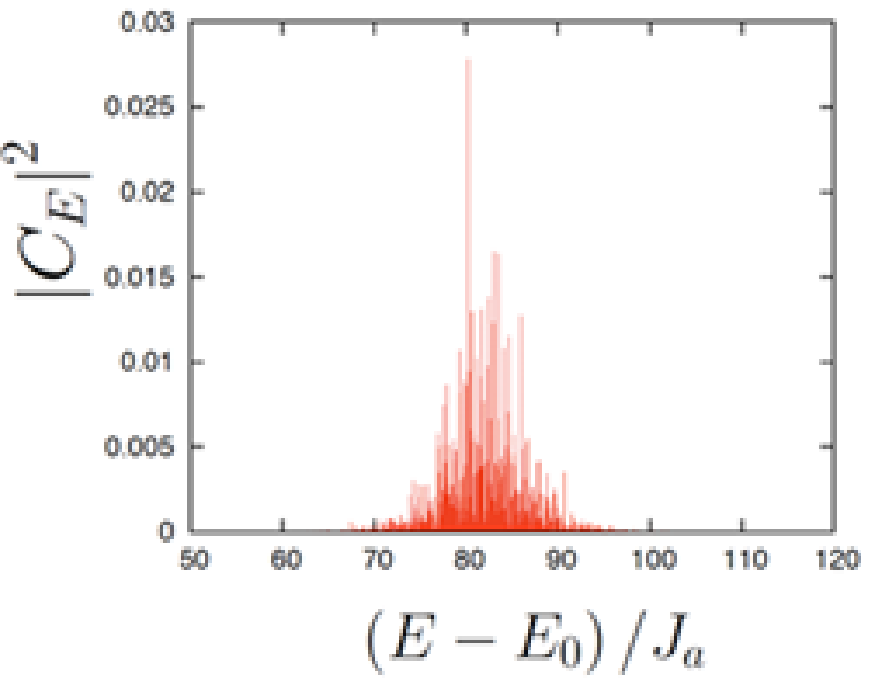}
\caption{ (color online) The distribution in the energy-eigenstates in the initial states ``{\it case1} ". This data is in the chaotic region $\sqrt{N} g / J_{a} = 4$. $E_{0}$ is the energy of the ground state.}
\label{fig:C_E_chaos_1}
\end{minipage}\hspace{1.0pc}
\begin{minipage}{18pc}
\includegraphics[width=0.9\linewidth, keepaspectratio]{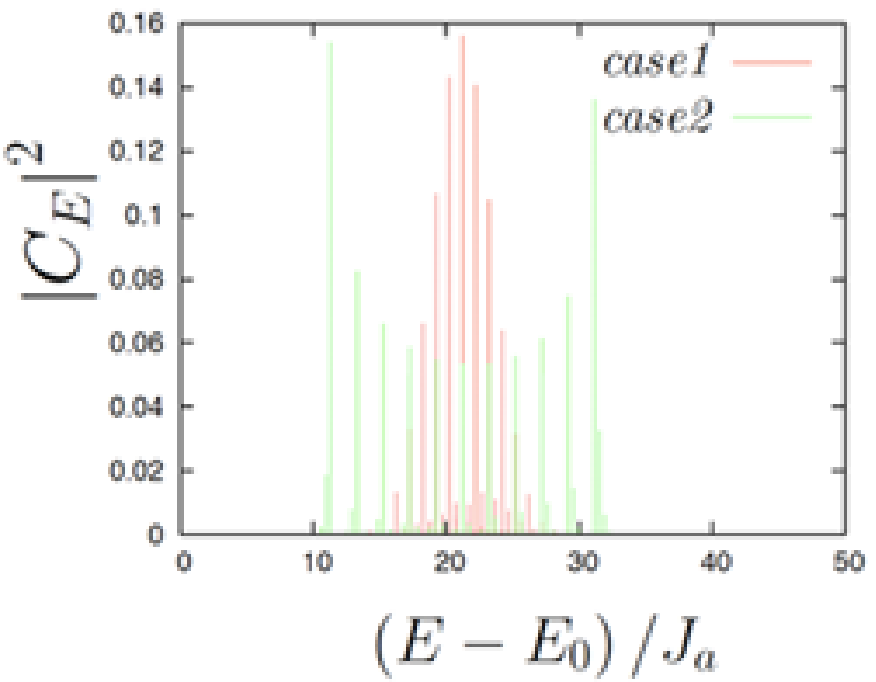}
\caption{ (color online) The distribution in the energy-eigenstates in the initial states ``{\it case1} " and ``{\it case2} ". The red represents ``{\it case1} ", and the green line is ``{\it case2} ". This data is in the regular region $\sqrt{N} g / J_{a} = 0.2$. $E_{0}$ is the energy of the ground state.}
\label{fig:C_E_R}
\end{minipage}
\end{figure}

\begin{figure}
\begin{center}
\includegraphics[width=7cm]{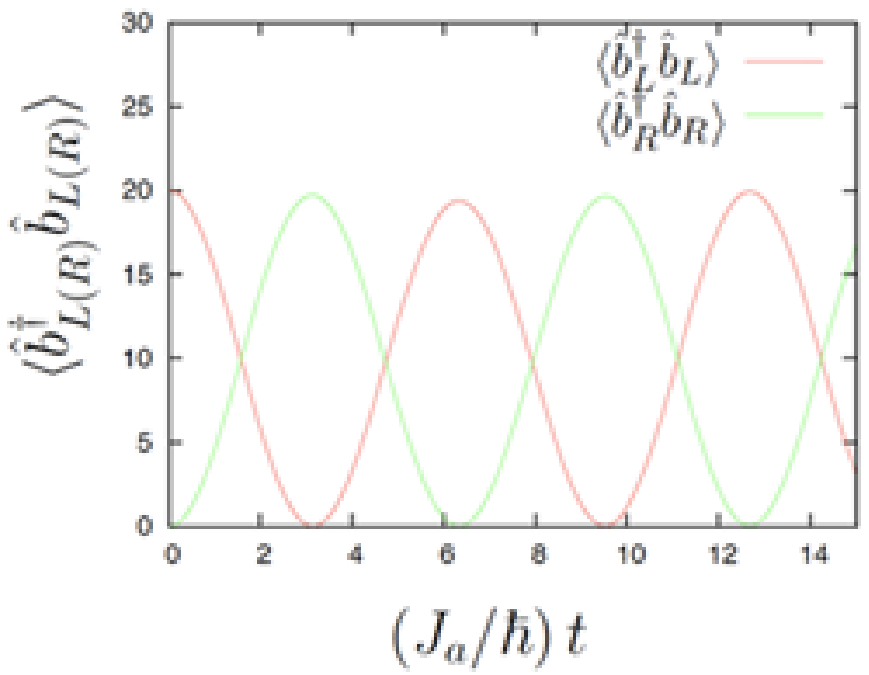}
\caption{ (color online) The initial state is ``{\it case1} ". This data is in the regular region $\sqrt{N} g / J_{a} = 0.2$. The microcanonical expectation value is $ \langle \hat{ b }_{ L \left( R \right) }^{ \dag } \hat{ b }_{ L \left( R \right) } \rangle^{ micro } \simeq 6$ in this parameter. }
\label{fig:R_dynamics_1}
\end{center}
\end{figure}

In Figs. \ref{fig:Time_evo_1_a}, \ref{fig:Time_evo_1_b}, \ref{fig:Time_evo_2_a} and \ref{fig:Time_evo_2_b}, we investigate the time-evolution in the chaotic region ($\sqrt{N} g / J_{a} = 4$), and compare the results from the different initial states {\it case1} and {\it case2}. As clearly shown in these figures, the dynamics is irreversible, i.e. the relaxation dynamics occurs. Furthermore, the relaxed values are the same in both the {\it case1} and {\it case2}. These relaxed values coincide with the microcanonical averages calculated by Eq. (\ref{eq:micro_canonical_mean_value}). When we calculate the microcanonical averages in this section, we set as $\Delta N = 20$.

In contrast to these relaxation dynamics, the dynamics are periodic in the regular region $\sqrt{N} g / J_{a} = 0.2$. The time evolutions of the molecular particle numbers are shown in Fig. \ref{fig:R_dynamics_1}. In this figure, the dynamics is periodic, and the relaxation to the microcanonical average does not occur. 

In the last of this section, we note that relaxation is induced by the dephasing process as
\begin{eqnarray}
\langle \hat{A} \rangle =  \sum_{\alpha, \beta} C_{\alpha}^{*} C_{\beta} e^{ - i \left( E_{ \alpha } - E_{ \beta } \right) t } A_{ \alpha \beta } \simeq \sum_{\alpha} \left| C_{\alpha} \right|^2 A_{\alpha, \alpha}. 
\label{eq:dephasing}
\end{eqnarray}
In the Wigner region, the distribution in the energy-eigenstates are spread out because of the level repulsion. As shown in Fig. \ref{fig:C_E_chaos_1}, the distribution in the energy eigenstates is close to be continuous. In this region, the many different period oscillations are superimposed, and the dephasing occurs. In contrast to this chaotic cases, in the Poisson region, the distribution in the energy eigenstates is discrete as shown in Fig. \ref{fig:C_E_R}. The energy levels are clustered such as $P_{P} \left( S \right) = e^{-S}$. In this region, the relaxation can not occur.

\subsection{Eigenstate thermalization hypothesis}
\label{sec:ETH}
\begin{figure}
\begin{minipage}{18pc}
\includegraphics[width=0.9\linewidth, keepaspectratio]{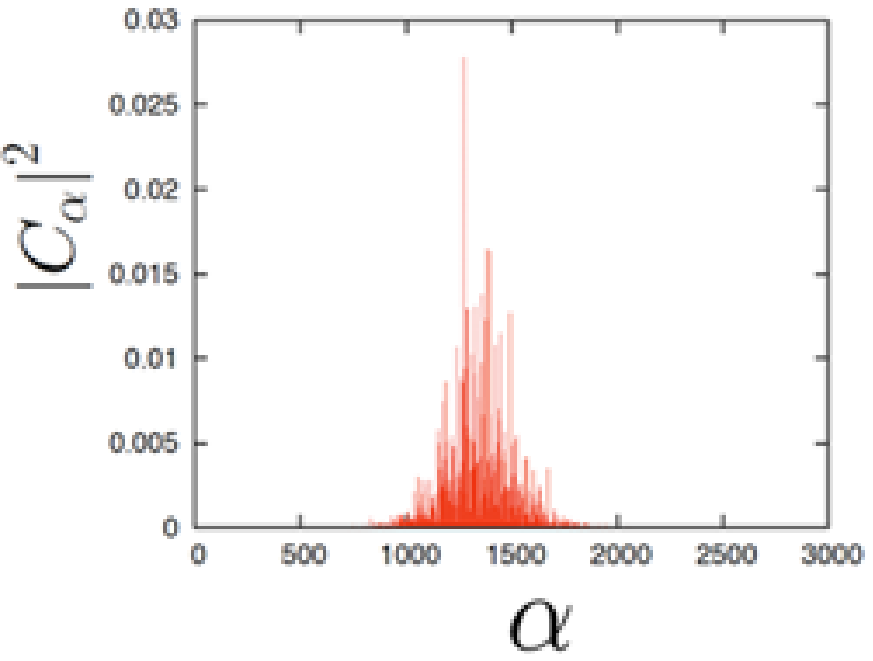}
\caption{ (color online) The distribution in the energy-eigenstates in the initial state ``{\it case1} ". This data is in the chaotic region $\sqrt{N} g / J_{a} = 4$. $\alpha$ is the numbering to the energy-eigenstate. The energy-eigenstates are arranged in ascending order.}
\label{fig:C_alpha_1}
\end{minipage}\hspace{1.0pc}
\begin{minipage}{18pc}
\includegraphics[width=0.9\linewidth, keepaspectratio]{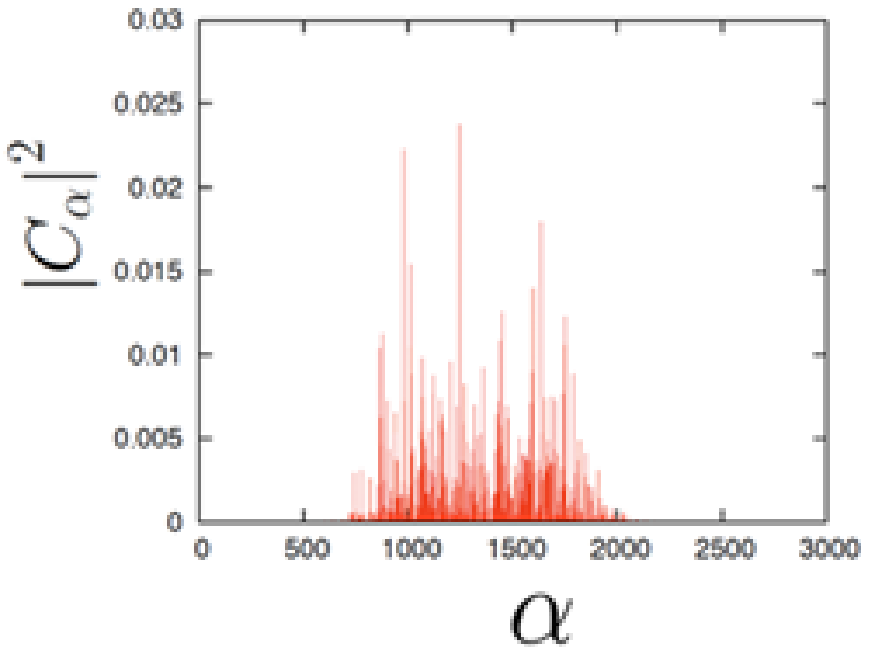}
\caption{ (color online) The distribution in the energy-eigenstates in the initial state ``{\it case2} ". This data is in the chaotic region $\sqrt{N} g / J_{a} = 4$. $\alpha$ is the numbering to the energy-eigenstate. The energy-eigenstates are arranged in ascending order.}
\label{fig:C_alpha_2}
\end{minipage}
\end{figure}

\begin{figure}
\begin{center}
\includegraphics[width=10.3cm]{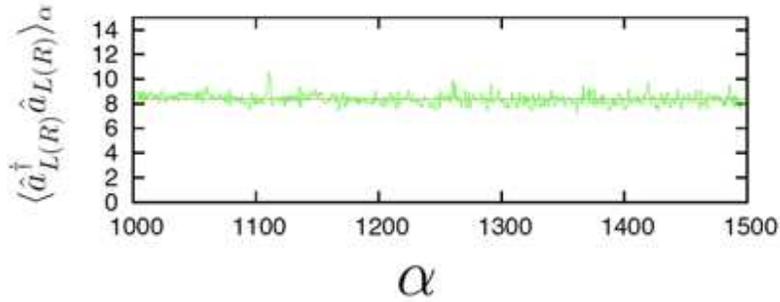}
\caption{ (color online) $\alpha$-dependence of the quantum-mechanical mean values in the chaotic region $\sqrt{N} g / J_{a} = 4$. $\alpha$ is the numbering to the energy-eigenstate. The energy-eigenstates are arranged in ascending order. This figure represents about the atomic-particle numbers in left and right wells. The straight line is the average of $ \langle \hat{ a }_{ L \left( R \right) }^{ \dag } \hat{ a }_{ L \left( R \right) } \rangle $. We set as $N=40$.}
\label{fig:ETH_chaotic_a}
\end{center}
\end{figure}

In this section, we verify the eigenstate thermalization hypothesis (ETH) \cite{Rigol_Nature}. In Figs. \ref{fig:Time_evo_1_a}, \ref{fig:Time_evo_1_b}, \ref{fig:Time_evo_2_a} and \ref{fig:Time_evo_2_b}. The relaxation to the same value occurs, although the distributions in the energy eigenstates are different in the initial condition ${\it case1}$ and ${\it case2}$, as shown in Figs. \ref{fig:C_alpha_1} and \ref{fig:C_alpha_2}. This is caused by the significant behavior of the energy eigenfunctions as shown in Fig. \ref{fig:ETH_chaotic_a}. In this figure, the quantum-mechanical mean values of physical quantities have the same value over many different energy eigenstates. This means that $A_{ \alpha, \alpha }$ is independent of $\alpha$ in (\ref{eq:dephasing}). This behavior of energy eigenfunctions is called the eigenstate thermalization hypothesis. ETH explains the thermalization mechanism in Figs. \ref{fig:Time_evo_1_a}, \ref{fig:Time_evo_1_b}, \ref{fig:Time_evo_2_a} and \ref{fig:Time_evo_2_b}.

Because of the independence of $A_{\alpha, \alpha}$ on the energy eigenstate labeling $\alpha$, the constant value of $A_{\alpha, \alpha}$ is almost the same as the microcanonical value defined in Eq. (\ref{eq:micro_canonical_mean_value}). We represent this value as $\langle A \rangle^{micro}$. Therefore the relaxed value is represented as
\begin{eqnarray}
\langle \hat{A} \rangle \simeq \sum_{\alpha} \left| C_{\alpha} \right|^2 A_{\alpha, \alpha} \simeq \langle A \rangle^{micro} \sum_{\alpha} \left| C_{\alpha} \right|^2 = \langle A \rangle^{micro}. \label{eq:alpha_independent}
\end{eqnarray}
 As explained above, when the eigenstate thermalization hypothesis is verified, the physical quantities relax to the microcanonical mean values, calculated by the principle of equal weight. It is clear from Eq. (\ref{eq:alpha_independent}) that the relaxed values are independent of the coefficient $C_{\alpha}$. Thermal equilibration is independent of the initial states. Therefore, the relaxation values are almost the same in the {\it case1} and {\it case2}, although the distributions in the energy-eigenstates are different as shown in Figs. \ref{fig:C_alpha_1} and \ref{fig:C_alpha_2}. In addition, it is clear from Fig. \ref{fig:ETH_chaotic_a} that the microcanonical ensemble average of atomic particle numbers defined as (\ref{eq:micro_canonical_mean_value}) do not depend on $\Delta N$. Moreover we note that the microcanonical ensemble averages of molecular particle numbers also satisfy ETH. Due to the total particle number conversation $N = 2 \left( \langle \hat{ a }_{L(R)}^{\dag} \hat{ a }_{L(R)} \rangle_{\alpha} + 2 \langle \Psi_{\alpha} | \hat{ b }_{L(R)}^{\dag} \hat{ b }_{L(R)} \rangle_{\alpha} \right)$, $\langle \hat{ b }_{L(R)}^{\dag} \hat{ b }_{L(R)} \rangle_{\alpha}$ satisfies ETH when $\langle \hat{ a }_{L(R)}^{\dag} \hat{ a }_{L(R)} \rangle_{\alpha}$ does ETH.

\begin{figure}
\begin{minipage}{18pc}
\includegraphics[width=0.9\linewidth, keepaspectratio]{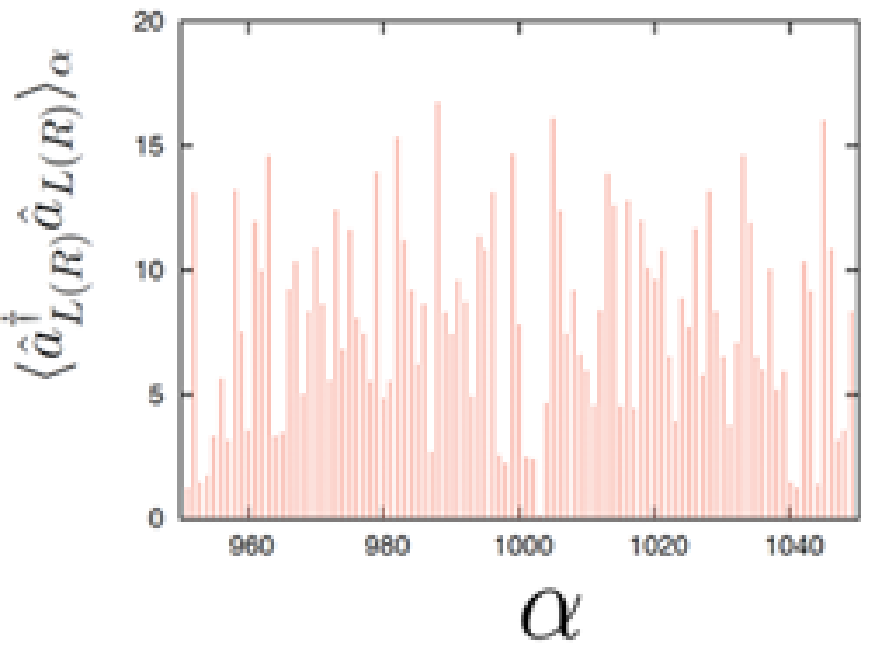}
\caption{ (color online) $\alpha$-dependence of the quantum-mechanical mean values in the regular region $\sqrt{N} g / J_{a} = 0.2$. $\alpha$ is the numbering to the enregy-eigenstate. In this figure, we focus on the atomic-particle numbers in the left and right wells.}
\label{fig:ETH_regular}
\end{minipage}\hspace{1.0pc}
\begin{minipage}{18pc}
\includegraphics[width=0.9\linewidth, keepaspectratio]{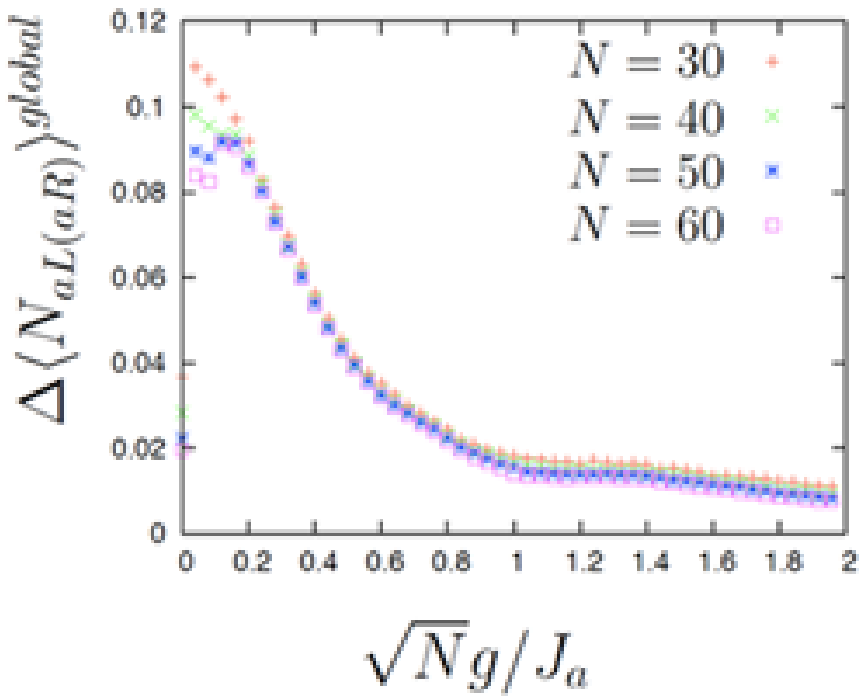}
\caption{ (color online) The fluctuations around the microcanonical mean values of the atomic-particle numbers in the left and right wells.}
\label{fig:ETH_chaos_a}
\end{minipage}
\end{figure}

In contrast to the chaotic cases, the quantum-mechanical mean values strongly depend on the energy eigenstate labeled by $\alpha$ in the regular region. This is clearly shown in Fig. \ref{fig:ETH_regular}. 

Next, we investigate whether chaos and ETH appear simultaneously. First, we evaluate fluctuations of quantum-mehcanical mean values around a microcanonical-mean value. By using the definition of Eq. (\ref{eq:micro_canonical_mean_value}), we represent the fluctuation around the $i$th energy eigenstate as 
\begin{eqnarray}
\Delta \langle \hat{a}^{\dag}_{L(R)} \hat{a}_{L(R)} \rangle^{micro}_{i} &\equiv& \frac{1}{N} \sqrt{ \frac{1}{2 \Delta N + 1} \sum_{ i^{'} = i - \Delta N}^{i + \Delta N} \left( \langle \hat{a}^{\dag}_{L(R)} \hat{a}_{L(R)} \rangle_{ i^{'} } - \langle \hat{a}^{\dag}_{L(R)} \hat{a}_{L(R)} \rangle^{micro}_{i} \right)^2 }, \label{eq:fluc_micro_a} \qquad \\
\Delta \langle \hat{b}^{\dag}_{L(R)} \hat{b}_{L(R)} \rangle^{micro}_{i} &\equiv& \frac{2}{N} \sqrt{ \frac{1}{2 \Delta N + 1} \sum_{ i^{'} = i - \Delta N}^{i + \Delta N} \left( \langle \hat{b}^{\dag}_{L(R)} \hat{b}_{L(R)} \rangle_{ i^{'} } - \langle \hat{b}^{\dag}_{L(R)} \hat{b}_{L(R)} \rangle^{micro}_{i} \right)^2 }, \qquad
\label{eq:fluc_micro_b}
\end{eqnarray}
where we set as $\Delta N = 10$. The maximum values of Eqs. (\ref{eq:fluc_micro_a}) and (\ref{eq:fluc_micro_b}) are normalized to 1. If Eqs. (\ref{eq:fluc_micro_a}) and (\ref{eq:fluc_micro_b}) are small, quantum-mehcanical mean values are almost the same, which means the realization of ETH. However, from Eqs. (\ref{eq:fluc_micro_a}) and (\ref{eq:fluc_micro_b}), we can evaluate the fluctuation around the $i$th energy eigenstate only. Therefore, we define the indicator of the verification of ETH as
\begin{eqnarray}
\Delta \langle N_{aL(aR)} \rangle^{global} \equiv \frac{1}{ N_{ETH} } \sum_{i=1, N_{ETH}} \Delta \langle \hat{a}^{\dag}_{L(R)} \hat{a}_{L(R)} \rangle^{micro}_{ \left[ i \times \left( 2 \Delta N + 1 \right) - \Delta N \right] },
\label{eq:global_fluc}
\end{eqnarray}
where $N_{ETH} \equiv \lfloor \left( E_{max} - E_{min} \right) / \left( 2 \Delta N +1 \right) \rfloor$. Here, $\lfloor x \rfloor$ is a floor function, which gives the largest integer less than $x$ or equal to $x$. Here we omit the lower and higher $20 \%$ energy eigenstates in the same way as Fig. \ref{fig:eta}. $E_{max}$ ($E_{min}$) is the maximum (minimum) value in this energy spectra. We numerically evaluate the verification of ETH by Eq. (\ref{eq:global_fluc}). We define the indicator $\Delta \langle N_{bL(bR)} \rangle^{global}$ for molecular particle number in the left (right) well in the same way as Eq. (\ref{eq:global_fluc}).

From Fig. \ref{fig:ETH_chaos_a}, ETH is verified with increasing the atom-molecule internal tunneling. In particular, the fluctuations about the microcanonical averages are suppressed when $\sqrt{N} g / J_{a} \ge 1$. $\Delta \langle N_{bL(bR)} \rangle^{global}$ exhibits exactly the same behavior as Fig. \ref{fig:ETH_chaos_a}. Comparing this result with Fig. \ref{fig:eta}, we conclude that the onset of chaos and ETH occurs simultaneously. 

\section{Particle-Localized Eigenstates}
\label{sec:Localized}
\begin{figure}
\begin{minipage}{18pc}
\includegraphics[width=0.9\linewidth, keepaspectratio]{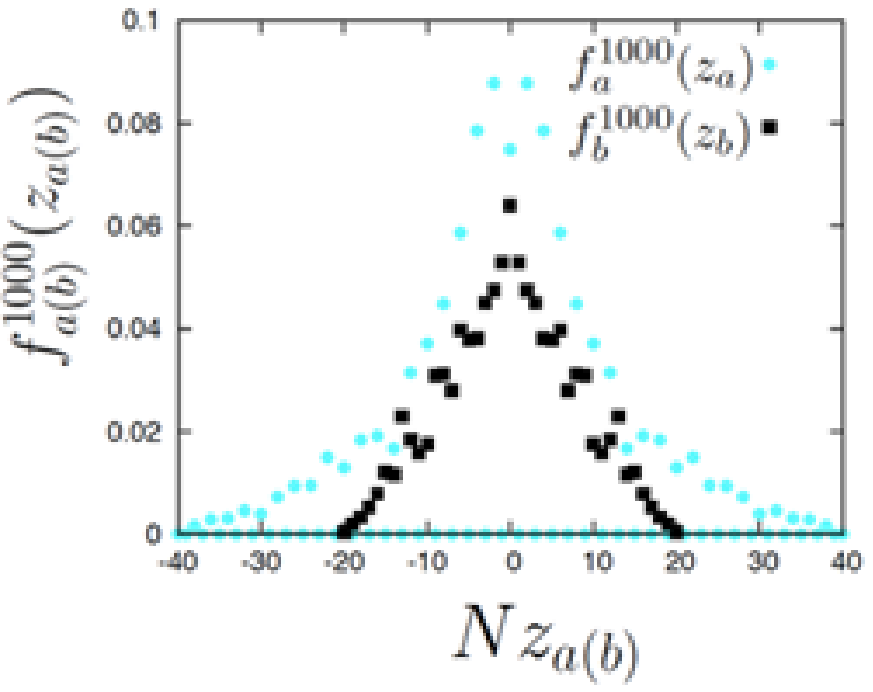}
\caption{ (color online) The distribution functions of $z_{a(b)}$. We investigate the 1000th energy eigenstate in the Wigner region $\sqrt{N} g / J_{a} = 4$.}
\label{fig:L_Za_Zb}
\end{minipage}\hspace{1.0pc}
\begin{minipage}{18pc}
\includegraphics[width=0.9\linewidth, keepaspectratio]{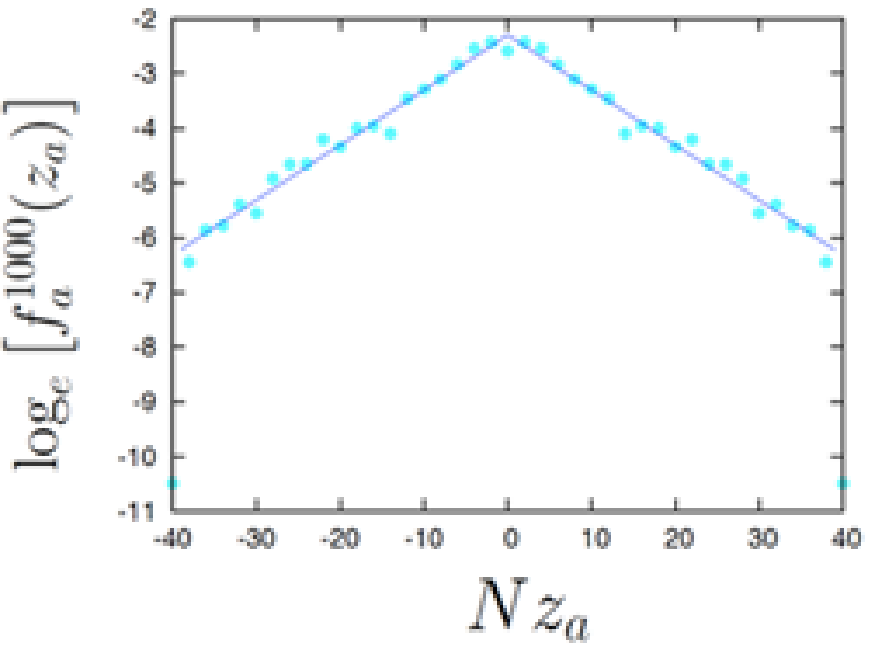}
\caption{ (color online) The logarithm of $f^{1000}_{a} ( z_{a} )$ in Fig. \ref{fig:L_Za_Zb}. We omit the odd numbers of $N z_{a}$. The straight line is the result of the least-square method.}
\label{fig:L_log}
\end{minipage}
\end{figure}

We show that the atom-molecule internal tunneling change energy eigenstates to be exponentially localized. We investigate the case $N=40$ throughout Sec. \ref{sec:Localized}.

We write the interwell relative atomic (molecular) particle number as $z_{a(b)} \equiv \left( N_{aL(bL)} - N_{aR(bR)} \right) / N$. Here we define the distribution function of the interwell relative atomic (molecular) particle as 
\begin{eqnarray}
f^{i}_{a(b)} \left( z_{a(b)} \right) \equiv \sum_{N_{aL}, N_{aR}, N_{bL}, N_{bR} } | \Phi_{i} \left( N_{aL}, N_{aR}, N_{bL}, N_{bR} \right) |^2 \delta_{N z_{a(b)}, N_{aL(bL)} - N_{aR(bR)} },
\label{eq:dis_Z_ab}
\end{eqnarray}
which is the distribution function of the $i$th energy eigenstate. 

In Fig. \ref{fig:L_Za_Zb}, we show the distribution functions of interwell relative particle numbers in the Wigner region $\sqrt{N} g / J_{a} = 4$. In this figure, we investigate the 1000th energy eigenstate. When $N z_{a}$ is an odd number, $f^{i}_{a} \left( N z_{a} \right)$ is zero because the total particle number $N$ is the even number. Furthermore, we can see the exponential localization of the distribution functions in both $f^{1000}_{a} \left( N z_{a} \right)$ and $f^{1000}_{b} \left( N z_{b} \right)$. In Fig. \ref{fig:L_log}, we plot the logarithm of $f_{a} \left( N z_{a} \right)$ of the 1000th energy eigenstate. We omit the points where $N z_{a}$ is an odd number. In this figure, by using the least-square method, we can approximate the distribution function by straight lines. When we plot the straight lines, we omitted the edge points such as $z_{a} = \pm 1$. Because of the finite size effect, these points are not approximated well by straight lines. 

\begin{figure}
\begin{minipage}{18pc}
\includegraphics[width=0.9\linewidth, keepaspectratio]{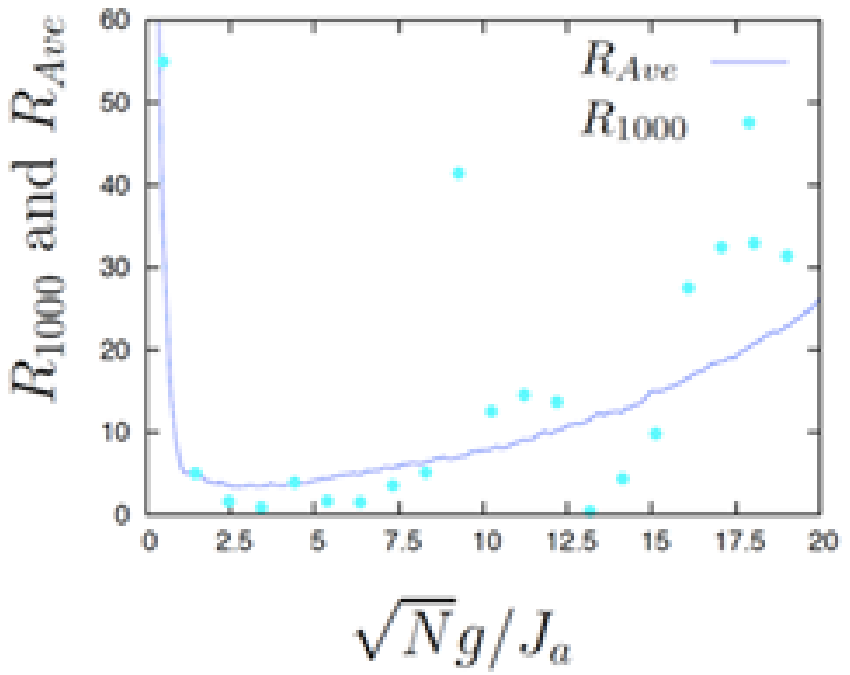}
\caption{ (color online) The difference between the energy eigenstates and exponent functions defined in Eq. (\ref{eq:dif_R}) and Eq. (\ref{eq:R_ave}).}
\label{fig:L_g_large}
\end{minipage}\hspace{1.0pc}
\begin{minipage}{18pc}
\includegraphics[width=0.9\linewidth, keepaspectratio]{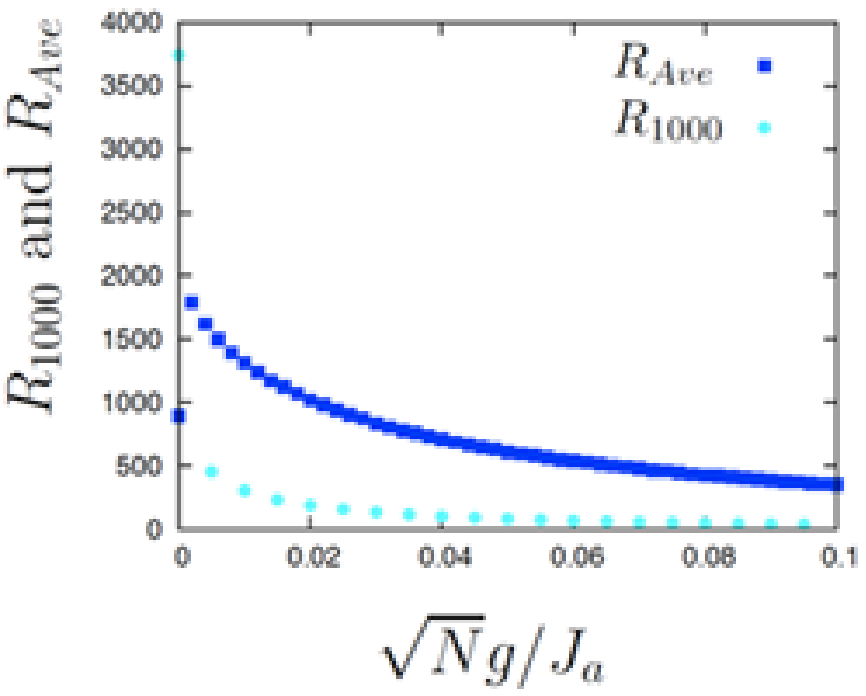}
\caption{ (color online) The difference between the energy eigenstates and exponent functions defined in Eq. (\ref{eq:dif_R}) and Eq. (\ref{eq:R_ave}). In this figure, we focus on the Poisson region ($0 \le \sqrt{N} g / J_{a} \le 0.1$).}
\label{fig:L_g_small}
\end{minipage}
\end{figure}

Next, we evaluate quantitatively the difference between the distribution function and the exponent functions. We evaluate this difference numerically as
\begin{eqnarray}
R_{i} \equiv \sum_{ N z_{a}=0}^{N-1} \left[ \textrm{log} \left[ f^{i}_{a} \left( N z_{a} \right) \right] - \left( C_{0} + C_{1} N z_{a} \right) \right]^2 ,
\label{eq:dif_R}
\end{eqnarray}
where we decide the coefficients $C_{0}$ and $C_{1}$ by using the least-squared method. When we calculate $\sum_{ N z_{a}=0}^{N-1}$, we omit odd numbers of $N z_{a}$. In Eq. (\ref{eq:dif_R}), we use the $i$th energy eigenstate. Furthermore, we define the average value over a number of the energy eigenstates. 
\begin{eqnarray}
R_{Ave} \equiv \sum_{i=1000}^{2000} \frac{ R_{i} }{ 1001 },
\label{eq:R_ave}
\end{eqnarray}
where we average the $R_{i}$s from 1000th to 2000th energy eigenstates. When $N=40$, the number of the energy eigenstates is $3311$. From 1000th to 2000th energy eigenstates exist in the middle of energy spectra. We plot the $R_{1000}$ and $R_{Ave}$ in Figs. \ref{fig:L_g_large} and \ref{fig:L_g_small}. 

Fig. \ref{fig:L_g_large} shows that the energy eigenstates behave exponentially when the interwell and internal tunneling compete. While $R_{1000}$ and $R_{Ave}$ are extremely large in the region $0 \le \sqrt{N} g / J_{a} \le 0.1$ as shown in Fig. \ref{fig:L_g_small}, $R_{1000}$ and $R_{Ave}$ are small in the region $\sqrt{N} g \sim J_{a}$ as shown in Fig. \ref{fig:L_g_large}. Comparing this behavior with Fig. \ref{fig:eta}, we conclude that the exponentially localization of the energy eigenstates occurs simultaneously with chaos. This means that the energy eigenstates behave as exponentially localized functions when thermalization occurs. In contrast, the energy eigenstates are quite different from exponential functions in the regular region as shown clearly in Fig. \ref{fig:L_g_small}.

\begin{figure}
\begin{minipage}{18pc}
\includegraphics[width=0.9\linewidth, keepaspectratio]{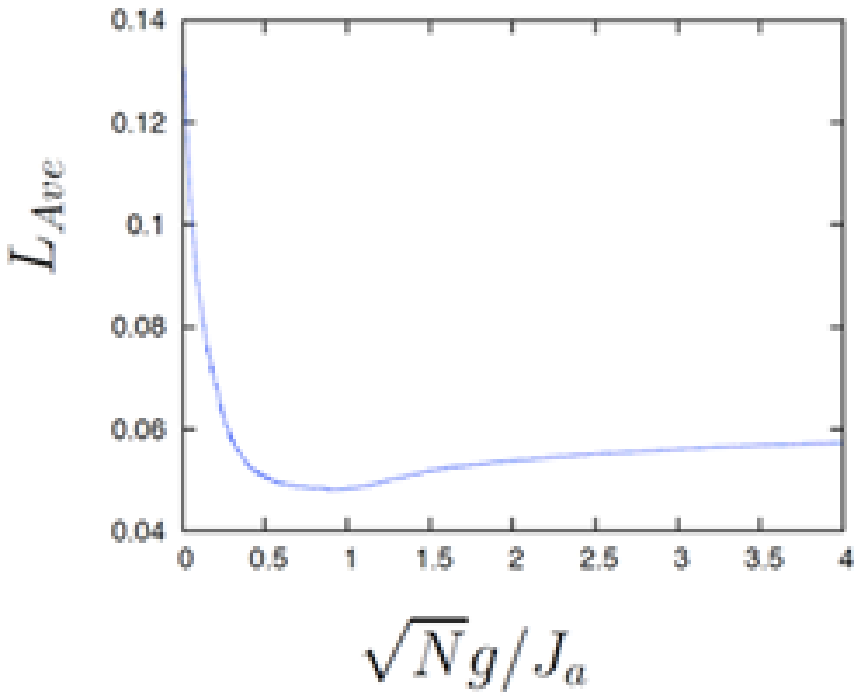}
\caption{ (color online) The indicator for the localization of energy eigenstates. $L_{Ave}$ is defined in Eq. (\ref{eq:L_Ave}).}
\label{fig:L_P}
\end{minipage}\hspace{1.0pc}
\begin{minipage}{18pc}
\includegraphics[width=0.9\linewidth, keepaspectratio]{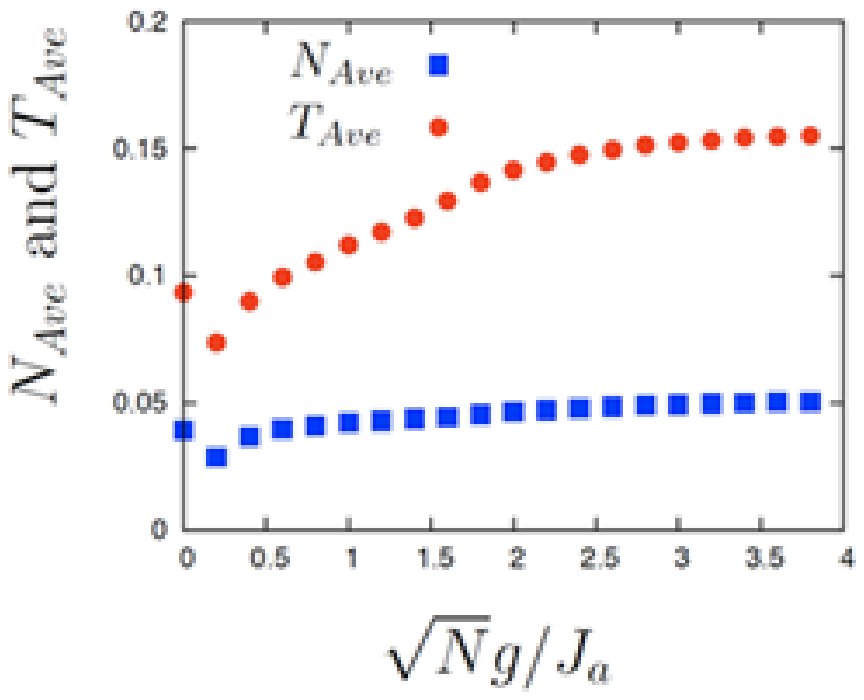}
\caption{ (color online) The fluctuations $N_{Ave}$ and $T_{Ave}$ defined in Eqs. (\ref{eq:N_Ave}) and (\ref{eq:T_Ave}).}
\label{fig:P_F}
\end{minipage}
\end{figure}

Next, we compare $R_{1000}$ with $R_{Ave}$. In both Figs. \ref{fig:L_g_large} and \ref{fig:L_g_small}, $R_{1000}$ with $R_{Ave}$ do not coincide. In particular, the 1000th energy eigenstate does not always localize exponentially as shown in Fig. \ref{fig:L_g_large}. $R_{1000}$ fluctuates. The exponential localization clearly appears in the averaged value $R_{Ave}$. This fact indicates that {\it the significant nature of the chaotic eigenstates manifests itself by some kind of statistical process}. This is the same as the idea of the level statistics. 

Next we show that the distribution functions are delocalized when the energy eigenstates behave as exponential functions. We define the indicator for localization as
\begin{eqnarray}
L_{i} \equiv \sum_{N z_{a} = - N}^{N} \left[ f^{i}_{a} \left( N z_{a} \right) \right]^2.
\end{eqnarray}
When the distribution function of the $i$th energy eigenstate $f^{i}_{a} \left( N z_{a} \right)$ is localized in a specified $N z_{a}$ only, $L_{i}$ is unity. When the distribution function of the $i$th energy eigenstate $f^{i}_{a} \left( N z_{a} \right)$ is spread out over many $N z_{a}$, $L_{i}$ is close to zero. Furthermore we define the average of $L_{i}$s over many energy eigenstates as
\begin{eqnarray}
L_{Ave} \equiv \sum_{1000}^{2000} \frac{ L_{i} }{1001}.
\label{eq:L_Ave}
\end{eqnarray}
We investigate $L_{Ave}$ in Fig. \ref{fig:L_P}. From this figure, delocalization occurs in the chaotic region. In contrast energy eigenfunctions are localized in the regular region. 

In Figs. \ref{fig:L_g_large} and \ref{fig:L_g_small}, we investigated the exponential localization. By comparing this with Fig. \ref{fig:L_P}, we note that the energy eigenfunctions are {\it delocalized} when the exponential localization occurs. When thermalization occurs, the particles move freely.

In the last of this section, we investigate the interwell atomic phase fluctuations. The canonical-conjugate variable of the interwell relative particle number $z_{a(b)}$ is the interwell relative phase \cite{Motohashi_P3}. When $z_{a(b)}$ behaves as exponential functions, the interwell phase fluctuation is increased as shown below. 

We define the fluctuation of the interwell atomic tunneling in the $i$th energy eigenstate as 
\begin{eqnarray}
T_{i} \equiv \frac{1}{N} \sqrt{ \left| \Bigg{\langle} \left( \frac{ \hat{a}^{\dag}_{L} \hat{a}_{R} + \hat{a}^{\dag}_{L} \hat{a}_{R} }{2} \right)^2 \Bigg{\rangle_{i}} - \left( \Bigg{\langle} \frac{ \hat{a}^{\dag}_{L} \hat{a}_{R} + \hat{a}^{\dag}_{L} \hat{a}_{R} }{2} \Bigg{\rangle_{i}} \right)^2 \right| },
\end{eqnarray}
where the maximum value is unity. We define the average of $T_{i}$s over many energy eigenstates as
\begin{eqnarray}
T_{Ave} \equiv \sum_{601}^{2600} \frac{ T_{i} }{2000}.
\label{eq:T_Ave}
\end{eqnarray}
Furthermore we define the fluctuation of $\hat{a}^{\dag}_{L} \hat{a}_{L} \hat{a}^{\dag}_{R} \hat{a}_{R}$ as
\begin{eqnarray}
N_{i} \equiv \frac{1}{N} \sqrt{ \Big{\langle} \left( \hat{a}^{\dag}_{L} \hat{a}_{L} \hat{a}^{\dag}_{R} \hat{a}_{R} \right)^2 \Big{\rangle_{i}} - \left( \Big{\langle} \hat{a}^{\dag}_{L} \hat{a}_{L} \hat{a}^{\dag}_{R} \hat{a}_{R}  \Big{\rangle_{i}} \right)^2 },
\end{eqnarray}
where the maximum value is unity. We define the average of $N_{i}$s over many energy eigenstates as
\begin{eqnarray}
N_{Ave} \equiv \sum_{601}^{2600} \frac{ N_{i} }{2000}.
\label{eq:N_Ave}
\end{eqnarray}
In Fig. \ref{fig:P_F}, we investigate $T_{Ave}$ and $N_{Ave}$. When we replace the creation-annihilation operators by c-numbers as $\hat{a}_{L(R)} \simeq \sqrt{N_{aL(aR)}} e^{i \theta_{ aL(aR) } }$, $\left( \hat{a}^{\dag}_{L} \hat{a}_{R} + \hat{a}^{\dag}_{L} \hat{a}_{R} \right)/2$ becomes 
\begin{eqnarray}
\sqrt{N_{aL} N_{aR}} \cos \left( \theta_{aL} - \theta_{aR} \right).
\label{eq:Tunneling}
\end{eqnarray}
In Fig. \ref{fig:P_F}, $T_{Ave}$ is increased while $N_{Ave}$ saturates. From Eq. (\ref{eq:Tunneling}), this means that the interwell atomic phase fluctuation is increased. The exponential behavior of $z_{a}$ is accompanied with the increase of the interwell atomic phase fluctuation.

\section{The influence of the interparticle interaction}
\label{sec:Ua}
In this section, we instigate the influence of the interatomic interaction on our results. Throughout this section, we set $\sqrt{N} g / J_{a} = 2$. As shown in Fig. \ref{fig:eta}, chaos occurs at $\sqrt{N} g / J_{a} = 2$ when $N U_{a} / J_{a} = 0$. Furthermore, as shown in Fig. \ref{fig:ETH_chaos_a}, ETH is verified simultaneously. We investigate the influence of the finite $U_{a}$ on chaos and thermalization. 

\begin{figure}
\begin{center}
\begin{minipage}{18pc}
\includegraphics[width=0.9\linewidth, keepaspectratio]{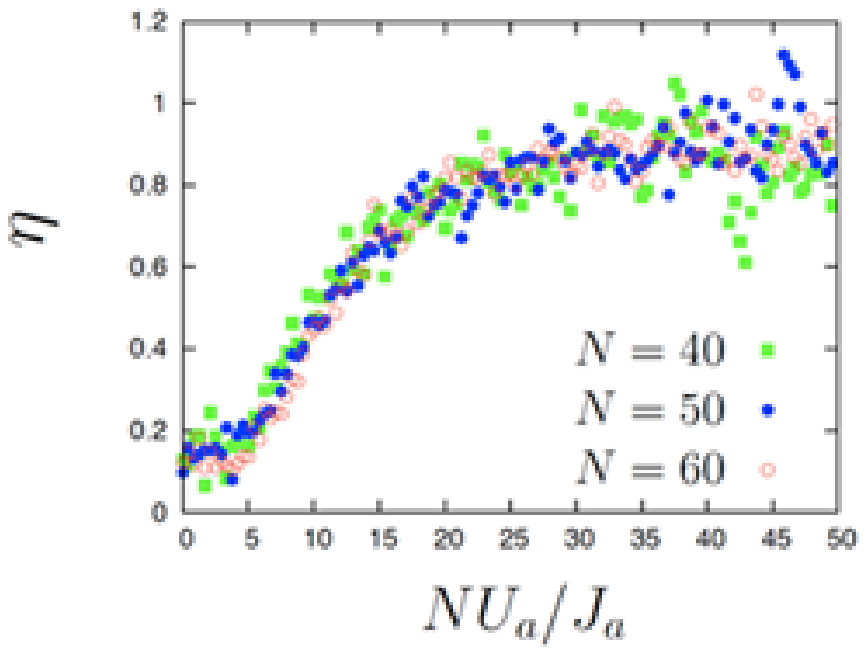}
\caption{ (color online) The $U_{a}$-dependence of $\eta$. In this figure, only even spectra are included. The lower and higher 20 \% levels are not included. We set $\sqrt{N} g / J_{a} =2$.}
\label{fig:eta_Ua}
\end{minipage}\hspace{0.5pc}
\begin{minipage}{18pc}
\includegraphics[width=0.9\linewidth, keepaspectratio]{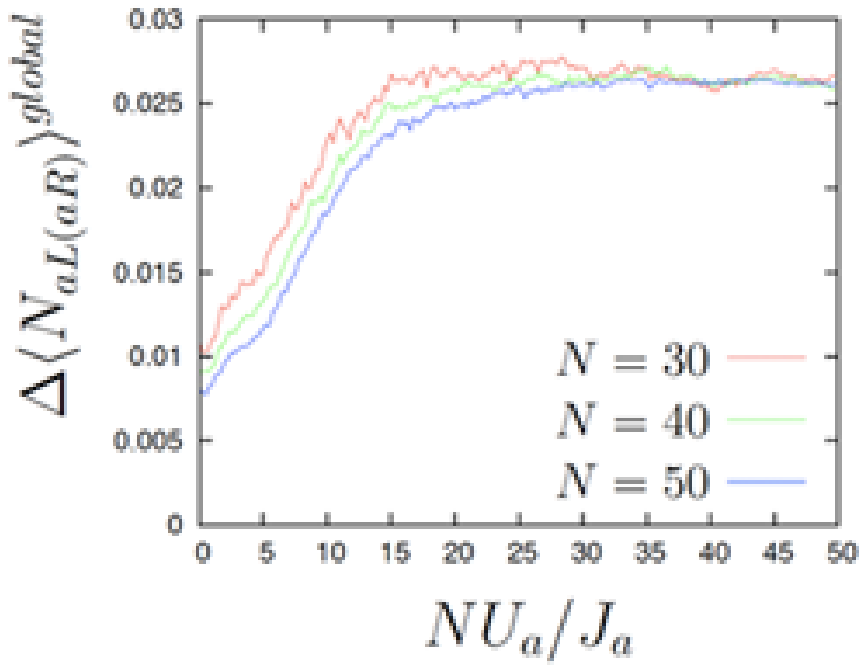}
\caption{ (color online) The fluctuations around the microcanonical mean values of the atomic-particle numbers in the left and right wells. We set $\sqrt{N} g / J_{a} =2$.}
\label{fig:ETH_Ua_A}
\end{minipage}
\end{center}
\end{figure}

In Fig. \ref{fig:eta_Ua}, we plot $\eta$ by increasing $N U_{a} / J_{a}$. In this figure we omit the lower and higher $20 \%$ energy eigenstates as in Fig. \ref{fig:eta}. From this figure, we conclude that the interatomic interaction suppresses chaos in our system. When $N U_{a} \sim J_{a}$, $\eta$ is near zero. The system is chaotic. However, $\eta$ is increased by the inter atomic interaction in the large $N U_{a} / J_{a}$ region. The periodicity of dynamics is recovered. When $N U_{a} / J_{a}$ is increased further, $\eta$ saturates to some constant below unity. This means that the periodicity of dynamics is not completely recovered.   

In Fig. \ref{fig:ETH_Ua_A}, we investigate the influence of the $U_{a}$ on ETH. In Fig. \ref{fig:ETH_Ua_A}, we investigate $\Delta \langle N_{aL(aR)} \rangle^{global}$. As same as $\eta$, $\Delta \langle N_{aL(aR)} \rangle^{global}$ is increased and saturates in the large $N U_{a} / J_{a}$ region. This indicates that the breakdown of ETH. However, comparing with Fig. \ref{fig:ETH_chaos_a}, the fluctuations $\Delta \langle N_{aL(aR)} \rangle^{global}$ are small in Figs. \ref{fig:ETH_Ua_A}. This reflects the fact that the revival of the periodicity is not complete. We note that $\Delta \langle N_{bL(bR)} \rangle^{global}$ exhibits exactly the same behavior as $\Delta \langle N_{aL(aR)} \rangle^{global}$.

\section{Summary}
\label{sec:summary}

We investigated how the atom-molecule internal tunneling deform the level-spacing distribution and dynamics. Increasing the internal tunneling strength, the level-spacing distribution changes to Wigner distribution from Poisson distribution. Further increasing the internal tunneling, the level spacing distribution becomes Poisson distribution again. The atom-molecule internal tunneling changes the dynamics significantly.

Furthermore, we investigated the thermalization dynamics. This thermalization is accompanied by the manifestation of eigenstate thermalization hypothesis (ETH) \cite{Rigol_Nature}. As explained in Sec. \ref{sec:ETH}, ETH explains the principle of equal weight in statistical mechanics from the microscopic view point. The foundation of statistical mechanics is based on the significant change of energy eigenstates. We showed numerically that the onset of ETH occurs simultaneously with that of chaos. 

We also found that the energy eigenstates become exponentially localized as shown in Sec. \ref{sec:Localized}. This exponential localization is analogous to Anderson localization. However, our result is significantly different from it in one respect. In the previous studies\cite{Cuevas_PRL,KR_2008,KR_2009_PRA,KR_1984,Garcia_PRE_2009}, the localization occurs in the insulator region, where the level statistics obeys the Poisson distribution. In contrast, the exponential localization appears in the Wigner region in our study. The opposite behavior occurs. 

In our study, interparticle interactions are not necessarily needed for thermalization. The atom-molecule internal tunneling induces thermalization in the absence of interparticle interactions. This point is different from the previous studies \cite{Rigol_Nature,Rigol_gapped,Kollath_Roux}. In these studies, Wigner distribution is induced in the presence of interparticle interaction. In Sec. \ref{sec:Ua} only, we include the interparticle interactions in our analysis. As a result, we showed that thermalization occurs when $N U_{a}$, $J_{a}$ and $\sqrt{N} g$ competes. It seems that the competition between parameters is important for thermalization.

We note that the multiplicity of degrees of freedom causes thermalization independently of integrability in general systems \cite{Sugita_2007}. When the number of degrees of freedom is large, thermalization occurs. This is quite different from ETH. Not the multiplicity of degrees of freedom but non-integrability plays an important role in ETH. At least, large system size is not needed for ETH. In this paper we show that ETH is verified in the system with only two sites. Our study clearly shows that large system size is not important for ETH. 

Finally, we note that the same result as atom-molecule Bose gases cannot occur in a binary Bose gases. By replacing the internal tunneling term in Eq. (\ref{eq:am_H_2}) with $\big( b_{L}^{\dag} a_{L} + b_{R}^{\dag} a_{R} + a_{L}^{\dag} b_{L} + a_{R}^{\dag} b_{R} \big)$, we obtain the Hamiltonian for a binary BEC mixture. From the same procedure as in this paper, we can calculate the level statistics of a binary Bose gases. As a result, it turned out that Wigner distribution does not appear. Thermalization is caused by atom-molecule internal tunneling. 

\newpage

\begin{acknowledgments}
The author would like to thank A. Sugita, H. Abuki, T. Monnai, T. Yamamoto, T. Nikuni and A. Watanabe for valuable comments and discussions..
\end{acknowledgments}

\appendix
\section{Parameters}
\label{sec:parameter_appendix}
The parameters in the four-mode Hamiltonian (\ref{eq:am_H_2}) are defined as follows :
\begin{eqnarray}
&& J_{i} \equiv - \int d \mathbf{r} \Phi_{iL}^{*} \left[ - \frac{\hbar^2}{ 2m_{i} } \nabla^{2} + V_{ \rm{ext} } ( \mathbf{r} ) \right] \Phi_{iR} ,
\label{eq:p_1} \\
&& E_{i}^{0} \equiv  \int d \mathbf{r} \Phi_{iL}^{*} \left[ - \frac{\hbar^2}{ 2m_{i} } \nabla^{2} + V_{ \rm{ext} } ( \mathbf{r} ) \right] \Phi_{iL} =  \int d \mathbf{r} \Phi_{iR}^{*} \left[ - \frac{\hbar^2}{ 2m_{i} } \nabla^{2} + V_{ \rm{ext} } ( \mathbf{r} ) \right] \Phi_{iR} ,
\label{eq:p_2} \\
&& U_{a} \equiv g_{a} \int d \mathbf{r} | \Phi_{iL} |^{4} = g_{a} \int d \mathbf{r} | \Phi_{iR} |^{4} ,
\label{eq:p_3} \\
&& g \equiv \lambda \int d \mathbf{r} \Phi_{bL}^{*} \Phi_{aL} \Phi_{aL} = \lambda \int d \mathbf{r} \Phi_{bR}^{*} \Phi_{aR} \Phi_{aR} ,
\label{eq:p_5} \\
&& \Delta \equiv \delta \int d \mathbf{r} | \Phi_{bL} |^{2} + E_{b}^{0} - 2 E_{a}^{0} = \delta \int d \mathbf{r} | \Phi_{bR} |^{2} + E_{b}^{0} - 2 E_{a}^{0} ,
\label{eq:p_6}
\end{eqnarray}
where $i = a$ $(b)$ represents atomic (molecular) lowest-energy modes, and L (R) expresses the left (right) well respectively.

\newpage

\bibliography{bunken}




\end{document}